\documentclass[10pt,twocolumn,letterpaper]{article}
\usepackage{ifpdf}
\ifpdf
  \pdfoutput=1
\fi

\usepackage[pagenumbers]{iccv} 

\definecolor{iccvblue}{rgb}{0.21,0.49,0.74}
\usepackage[pagebackref,breaklinks,colorlinks,allcolors=iccvblue]{hyperref}
\usepackage{multirow}
\usepackage{pifont}
\usepackage{tabularx}
\usepackage{colortbl}
\usepackage{afterpage} 
\usepackage{placeins}

\usepackage{stfloats} 

\def\paperID{4423} 
\def\confName{ICCV}
\def\confYear{2025}

\title{Proxy Prompt: Endowing SAM \& SAM 2 with Auto-Interactive-Prompt for Medical Segmentation}

\author{WANG XINYI\\
The Hong Kong Polytechnic University\\
{\tt\small haylee-xinyi.wang@connect.polyu.hk}
\and
KANG HONGYU\\
The Hong Kong Polytechnic University\\
{\tt\small hong-yu.kang@connect.polyu.hk}
\and
WEI PEISHAN\\
The Hong Kong Polytechnic University\\
{\tt\small peishan.wei@polyu.edu.hk}
\and
SHUAI LI\\
The Hong Kong Polytechnic University\\
{\tt\small sshuai.li@connect.polyu.hk}
\and
YU SUN\\
The Hong Kong Polytechnic University\\
{\tt\small sunxinbrier@foxmail.com}
\and
SAI KIT LAM\footnotemark[1]\\
The Hong Kong Polytechnic University\\
{\tt\small saikit.lam@polyu.edu.hk}
\and
YONGPING ZHENG\footnotemark[1]\\
The Hong Kong Polytechnic University\\
{\tt\small yongping.zheng@polyu.edu.hk}
}

\begin{document}
\maketitle
\begin{abstract}
In this paper, we aim to address the unmet demand for automated prompting and enhanced human-model interactions of SAM and SAM2 for the sake of promoting their widespread clinical adoption. Specifically, we propose Proxy Prompt (PP), auto-generated by leveraging non-target data with a pre-annotated mask. We devise a novel 3-step context-selection strategy for adaptively selecting the most representative contextual information from non-target data via vision mamba and selective maps, empowering the guiding capability of non-target image-mask pairs for segmentation on target image/video data. To reinforce human-model interactions in PP, we further propose a contextual colorization module via a dual-reverse cross-attention to enhance interactions between target features and contextual-embedding with amplifying distinctive features of user-defined object(s). Via extensive evaluations, our method achieves state-of-the-art performance on five public datasets and yields comparable results with fully-trained models, even when trained with only 16 image masks. 
\end{abstract}    
\section{Introduction}
\label{sec:intro}

Under the paradigm shifts in Large-scale Vision Models (LVMs), Segment Anything Model (SAM)~\cite{kirillov2023segment} and SAM 2~\cite{ravi2024sam} have been introduced as generalized foundation models for segmenting and tracking any objects on image and video data, respectively. These models provide a certain degree of interactive segmentation capacity as users can segment any target object(s) according to their needs by using a single model in-one-go. Such capabilities are achieved by leveraging the concept of prompts~\cite{wang2023review}, such as points, boxes, or masks, waiving the traditional demand for massive manually annotations. Instead, users are only required to input prompts directly on target images or video frames.

\begin{figure}[t]
  \centering
  \setlength{\abovecaptionskip}{3pt}
   \includegraphics[width=0.99\linewidth]{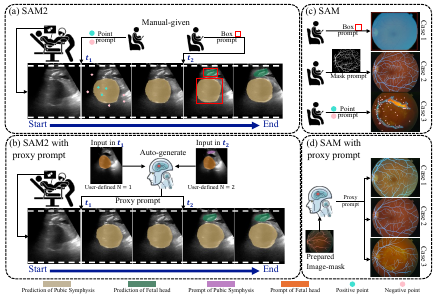}
   \caption{Illustration of comparison without/with PP in (a-b) SAM 2 using real-time ultrasound frames of 1 subject; and (c-d) SAM using a fundus retina dataset of 3 subjects.}
   \label{fig:Figure1}
   \vspace{-15pt}
\end{figure}

Notwithstanding, widespread clinical adoptions of these LVMs have been substantially impeded by the soaring medical demands for ``automated prompting'' and ``high-level human-model interactions'' when it comes to downstream medical tasks, particularly real-time imaging-guided interventions. The current prompting strategies are sub-optimal for two key reasons. First, medical image/video data entails an overwhelmingly huge variations in terms of complexity of the target object(s) to be segmented; segmenting such structures (\eg vessels) using existing prompts can be practically challenging. As illustrated in \cref{fig:Figure1}(c), for instance, manually inputting either box prompt (Case 1) or point prompt (Case 3) generates poor results; while adopting mask prompt (Case 2) performs well, it is highly tedious, exhaustive and knowledge-demanding for precise mask prompt formation. Second, users are required to input prompt for every single target image/video frame, which is a manual trial-and-error process, tedious, and not user-friendly. Clinical burden becomes exceedingly prominent when segmenting intricated structures \cref{fig:Figure1}(c) and/or massive datasets, especially in resource-limited clinics. Therefore, there is a pressing demand for automated prompt generation to accommodate various clinical needs.

\begin{figure}
  \centering
  \setlength{\abovecaptionskip}{3pt}
   \includegraphics[width=0.99\linewidth]{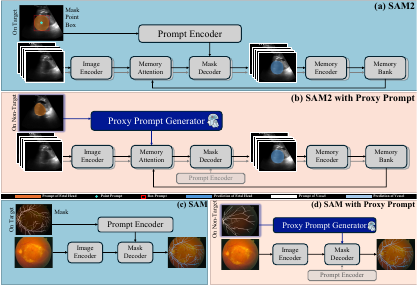}
   \caption{Schematic differences of traditional prompt encoder and our proposed PPG in SAM 2 (a-b) and SAM (c-d).}
   \label{fig:Figure2}
   \vspace{-15pt}
\end{figure}

Apart from this, the existing SAM-based models~\cite{shaharabany2023autosam, zhang2023customized, wu2023self} are inadequate to support the growing demand of high-level human-model interaction to accommodate multifarious clinical goals and high disparity in preferences of clinical users. These models segment target object(s) via training on specific single/multiple object(s). Yet, they do not adapt well to changes in the user’s preferences. For instance, in fundus imaging, optic dis/cup weights more for glaucoma detection~\cite{thakur2018survey}, while vessels are prioritized for assessing retinal vascular occlusion~\cite{sekhar2008automated}. Even though these objects may appear within the same image, switching segmentation task from the optic disc/cup to vessels necessitates retraining the model. In real-world clinics where tasks are greatly diverse, creating separate models for individual tasks is practically challenging and computational demanding. Therefore, it is imperative to reinforce human-model interaction capacity by allowing users to flexibly adjust prompts for satisfying various clinical demands without the need for model retraining. For example, in intrapartum ultrasound as illuminated in \cref{fig:Figure1}(a), these properties would provide high flexibility for users to segment fetal head (FH) alone or FH\&pubic symphysis (PS) at any time points \(\{t_{1}, t_{2}\}\) to achieve measurement of fetal head rotation or fetal position~\cite{ghi2018isuog}.

Confronted with these, we propose a novel Proxy Prompt (PP), which can be automatically generated from a ``non-target'' data (i.e., image/video frame of subjects other than the one under examination, such as from retrospective datasets) with a pre-annotated mask. This PP strategy is distinct from the existing prompting methods where prompting can only be made on ``target'' data, in a manual manner. As illustrated in \cref{fig:Figure1}(d), only one annotated image is required in using PP, tremendously streamlining workflow by waiving the prerequisite of providing separated prompts for every image/frame. Moreover, clinicians can freely switch segmentation tasks by adjusting the support-pair input anytime during examination without model retraining nor adopting different models, as shown in \cref{fig:Figure1}(b). Working in tandem with PP, we innovated a Proxy Prompt Generator (PPG) to reform SAM and SAM 2 for image and video data, respectively. Compared to SAM and SAM 2 in \cref{fig:Figure2} (a\&c), we employed high dimensional embedding from the PPG as prompts in \cref{fig:Figure2} (b\&d).

The core design of PPG lies in the novel Contextual Selective Module (CSM) and Contextual Colorization Module (CCM), which are dedicatedly configurated for automated prompting and high-level human-model interactions. First, CSM is introduced to enable adaptive selection of the most representative contextual information from ``non-target'' data for the ``target'' data, achieving cross-data guidance. Besides, CSM contains Vision Mamba, Bridge Unit, and Selective Map, implementing a 3-step selection process: (i) input-driven, (ii) object-guided, and (iii) target image/frame relevance selection to support both cross-video tasks and cross-image prompting. 
Second, CCM is devised to reinforce human-model interaction, enabling the model to interpret diverse user needs as indicated by different masks (\eg single/multiple objects). This aim is achieved by leveraging dual-reverse cross-attention to enhance the representation of contextual embedding. Finally, the PP, effectively capturing specific object features, is generated.
Such PPG-based strategy presents a simple yet efficient architecture to enhance clinical-friendliness of SAM and SAM 2, even in few-shot settings. Furthermore, with all original parameters frozen, our design can function as a flexible, plug-and-play module and can continuously adapt to the ever-evolving LVMs beyond SAM and SAM 2.

We conducted extensive experiments across several popular image and video datasets, validating the superior performance and stability of our proposed approach. Our main contributions are outlined below:

1. We propose a novel PP to enhance user-friendliness and precision of SAM and SAM 2 by equipping them with the capacity of automated prompting and high-level human-model interaction.

2. We devise CSM for adaptive selection of the most representative contextual information from “non-target” data to guide segmentation on “target” data, enabling effective cross-image/video prompting, waiving the need to execute prompting for every single target data, and minimizing experience-/expertise-derived variability in prompt quality.

3. We configurate CCM for enhancing the expressiveness of contextual embeddings to interpret and accommodate diverse clinical demands and preference of end users, thereby reinforcing model-human interactions.

4. Extensive experiments show that our model achieves state-of-the-art (SOTA) performance and is comparable to traditional segmentation models trained on full data volumes, even with only 16 image-mask pairs for SAM and SAM 2 training. Moreover, our strategy is of high potential to adapt to the iterative evolution of LVMs for medical tasks in the future.

\section{Related Work}
\label{sec:related}

\subsection{Adapting SAM to Medical Image Segmentation}
 Given the domain gap between natural and medical image datasets, various works~\cite{ma2024segment, wu2023medical, zhang2023customized, shaharabany2023autosam} have studied the application of SAM to medical image segmentation.  Medical SAM (MedSAM)~\cite{ma2024segment} fine-tunes the decoder of SAM using 1,570,263 medical image-mask pairs with bounding box prompts to adapt to medical tasks. Medical SAM Adapter (Med-SA)~\cite{wu2023medical} efficiently fine-tunes SAM using adapter with point and box prompts. SAM Medical (SAMed) model~\cite{zhang2023customized} efficiently fine-tunes the image encoder of SAM using another technique---low-rank-based (LoRA) strategy~\cite{hu2021lora}. AutoSAM~\cite{shaharabany2023autosam} utilizes the gradients provided by a frozen SAM to train a new encoder, thus automatically extracts prompts from the image itself to achieve automatic segmentation. However, these SAM-based works on medical data can be roughly categorized into two types: one inherits the prompt design of SAM but without the automatic capability (MedSAM and Med-SA); the other can automatically segment objects but sacrifices the human-model interaction (SAMed and AutoSAM).

\subsection{SAM without Manual Given Prompt}
In addition to SAMed and AutoSAM, many other works also focus on enhancing the automatic prompting capability of SAM to improve user-friendliness. Self-prompting SAM~\cite{wu2023self} fine-tune a self-prompt unit to first provide coarse segmentations, from which they extract point/box prompts for SAM to obtain the final results. Personalization approach for SAM (PerSAM) ~\cite{zhang2023personalize} identify the most similar point between the reference and test image as the prompt for SAM. Evidential prompt generation method (EviPrompt)~\cite{xu2023eviprompt} and ProtoSAM~\cite{Ayzenberg2024ProtoSAMOM} both adapt the idea in PerSAM into medical image domain. EviPrompt fits the point prompts according to image similarity, while ProtoSAM ultilizes reference image/mask pair to obtain a coarse segmentation of target image, then extracts point or box as prompts required by SAM.
However, these methods are still limited in simulating prompts such as points or boxes, which restricts their capability in the vessel-like branching structures due to the ambiguous instruction~\cite{ma2024segment}. In contrast, our PP emphasizes the high-level embeddings as prompts, thereby guiding the model with deeper-level information to precisely segment such intricate objects.

\subsection{Vision Backbone Based on Mamba}
To extract contextual information for guiding model segmentation, the recently designed Mamba can serve as a potential choice for building vision backbone. Based on the state space model (SSM)~\cite{gu2023modeling}, Mamba~\cite{gu2023mamba} boosts the development of SSM from the key aspects of ``structure''. In terms of structure, it breaks the input-invariant feature of the conventional SSM layer and constructs an input-dependent SSM layer, enabling it to focus on the effective information in the input. Various studies~\cite{ma2024fer,ju2024vm,zheng2024fd,ma2024u,zhu2024vision} have migrated Mamba to the vision domain. Facial Expression Recognition-YOLO-Mamba (FER-YOLO-Mamba)~\cite{ma2024fer} combines Mamba with attention to construct a dual-branch structure for facial expression detection. Vision Mamba-Denoising Diffusion Probabilistic Model (VM-DDPM)~\cite{ju2024vm} introduces Mamba in the medical image synthesis domain, utilizing an SSM-CNN hybrid structure within the diffusion model. Vision Mamba (Vim) solely relies on the SSM to construct a vision backbone, selectively capturing key information in the input-dependent manner, making it highly suitable for handling high-resolution inputs~\cite{zhu2024vision}, which is common in medical tasks. 
 \begin{figure*}
  \centering
  \setlength{\abovecaptionskip}{3pt}
   \includegraphics[width=0.9\linewidth]{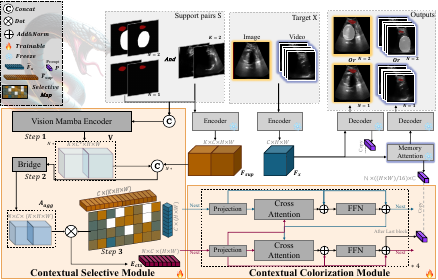}
   \caption{Designed Proxy Prompt Generator for both SAM 2 and SAM. Our key designed focus on the Contextual Selective Module and the Contextual Colorization Module. The Encoder and Decoder refer to the original structures, which are frozen.}
   \label{fig:Figure3}
   \vspace{-15pt}
\end{figure*}
\section{Method}
\label{sec:method}

Our proposed PPG, compatible with SAM and SAM 2, is illustrated in \cref{fig:Figure3}. Utilizing support image-mask pairs, this network not only enables auto-segmentation on target images or videos but also builds human-model interaction based on user-provided mask. For a target \(\mathbf{X}\), we define a support set \({S} = \{ {s}_i \}_{i=1}^N\) to assist in the segmentation of \(N\) objects within the target \(\mathbf{X}\). Each \({s}_i\) comprises \(K\) image and mask pairs denoted as \({s}_i = \{ (\mathbf{I}^j, \mathbf{M}_i^j) \}_{j=1}^K\), where each image \(\mathbf{I}^j \in \mathbb{R}^{3 \times H^0 \times W^0}\) and the corresponding mask \(\mathbf{M}_i^j \in \mathbb{R}^{H^0 \times W^0}\). First, the target \(\mathbf{X}\) and support images \(\{ \mathbf{I}^j \}_{j=1}^K\) are processed by a shared image encoder, with LoRA~\cite{hu2021lora} applied to adapt with medical tasks, resulting in \(\mathbf{F}_x \in \mathbb{R}^{ C \times H \times W}\) and \(\mathbf{F}_{\text{sup}} \in \mathbb{R}^{ K \times C \times H \times W}\). In parallel, our Contextual Selective Module (CSM) processes the support pairs through a three-layer selection mechanism to identify the most valuable contextual information for target segmentation, thereby enabling cross-frame prompting (\cref{sec:Selective}). Secondly, the contextual information \(\mathbf{E}_{\text{ctx}}\) is refined within the Contextual Colorization Module (CCM), ``coloring'' target features with user specifications via dual-reverse cross-attention to achieve human-model interaction (\cref{sec:Colorization}). Finally, this refined high-level embedding—enriched with user-defined information and object-specific target feature—serves as a high-dimensional prompt for SAM or SAM 2 (\cref{sec:SAM 2andSAM}). In summary, our method offers three main characteristics: cross-frame guidance, human-model interaction, and high-dimensional prompting via CSM and CCM. In the following sections, we detail each module step-by-step.

\subsection{Contextual Selective Module}
\label{sec:Selective}
We leverage Vision Mamba~\cite{zhu2024vision}’s selective input capabilities, establish communication between different objects, and design a filtering mechanism to refine the contextual embedding. As illustrated in \cref{fig:Figure3}, support pairs \(\{ {s}_i \}_{i=1}^N\) are initially processed by the Vision Mamba encoder to extract feature \(\mathbf{V}\) with input-adaptive parameters, marking the first selective step. Secondly, within the designed Bridge Unit, we enable inter-object communication across the concatenated features, thereby selectively extracting object-level feature \(\mathbf{A}_{\text{agg}}\). Thirdly, features (\(\mathbf{F}_{\text{x}}\)) and (\(\mathbf{F}_{\text{sup}}\)) are used to compute a \(\mathbf{Selective}\) map. Each row in this map representing a given patch from a support image will positively influence a specific patch in \( \mathbf{X} \). This selective map subsequently filters the aggregated information \(\mathbf{A}_{\text{agg}}\) to derive the most valuable information as the contextual embedding \(\mathbf{E}_{\text{ctx}}\). The following sections will detail each component of this pipeline.

{\bf First Selection Step: Vision Mamba.} We use the encoder of Vision Mamba (Vim) ~\cite{zhu2024vision} as the contextual encoder, as its input-dependent selection mechanism is well-suited for learning similar modality features, thereby enhancing prompt effectiveness. This mechanism enables parameters (i.e., \( \mathbf{\overline{A}} \), \( \mathbf{\overline{B}} \) and \(\mathbf{C}\))  influencing interactions within the image to adapt as functions of the input, allowing features to be extracted based on the input itself. Specifically, Vim first flattens the input 2-D image into patches and form the token sequence \( x \), which is progressively transformed into the output token sequence \( y \) according to the following formula, ultimately being encoded by us into the feature matrix.
\begin{equation}
  h_{t} = \mathbf{\overline{A}}h_{t-1} + \mathbf{\overline{B}}x_{t}
\end{equation}
\begin{equation}
  y_{t} = \mathbf{C}h_{t}
\end{equation}
where \( t \) represents the timestep and \( h \) denotes the latent state. The parameters \( \mathbf{\overline{A}} \), \( \mathbf{\overline{B}} \) and \(\mathbf{C}\) are generated depending on the input \( x \), rather than being input-invariant, thus allowing us to selectively encode the input image as our first step in the selection strategy. For more details, refer to~\cite{zhu2024vision}.

Given the support set \({S} = \{ {s}_i \}_{i=1}^N\), we fix a value \(K\) such that each \({s}_i = \{ (\mathbf{I}^j, \mathbf{M}_i^j) \}_{j=1}^K\) provides \(K\) support pairs \((\mathbf{I}^j, \mathbf{M}_i^j)\). For the \(j\)-th pair \(\{ (\mathbf{I}^j, \mathbf{M}_i^j) \}_{i=1}^N\) among the \(N\) support subset \({s}_i\), each masks \(\mathbf{M}_i^j\) correspond to one same image \(\mathbf{I}^j\). As a result, \(\mathbf{I}^j\) is concatenated with \(\mathbf{M}_i^j\) and fed into the contextual encoder, resulting in \(N\) feature matrices \(\mathbf{V} \in \mathbb{R}^{K \times C^{v} \times H \times W}\), where \(C^{v}\) indicates the channel of \(\mathbf{V}\). The process of extracting the feature matrix \(\mathbf{V}\) can be breifly represented by the following equation:
\begin{equation}
\mathbf{V} = \text{Vision\,Mamba\,Encoder}(\text{concat}(\mathbf{I}^j, \mathbf{M}_i^j))
\end{equation}

{\bf Second Selection Step: Bridge Unit.} Subsequently, in the Bridge Unit, we initially duplicate \(N\) times the obtained feature \(\mathbf{F}_{\text{sup}}\) and concat with \(\mathbf{V}\) of each object along the channel dimension to further aggregate the features. Most importantly, we employ convolutional block attention module(CBAM)~\cite{woo2018cbam} along the concatenated channel dimension to facilitate implicit inter-target communication, allowing the model to select key features across multiple objects, thereby enhancing cross-target contextual understanding. Additionally, two ResBlocks ~\cite{he2016deep} are used to prevent further feature dimension expansion. After the final block, the feature is flattened to form \(\mathbf{A}_{agg}\). This process is summarized as follows:
\begin{equation}
\mathbf{F}_{\text{cat}} = \text{concat}(\mathbf{F}_{\text{sup}}, \mathbf{V})
\end{equation}
\begin{equation}
\mathbf{A}_{agg} = \text{ResBlock}_2(\text{CBAM}(\text{ResBlock}_1(\mathbf{F}_{\text{cat}})))
\end{equation}
where  \(\mathbf{F}_{\text{cat}} \in \mathbb{R}^{K \times N \times (C^{v} + C) \times H \times W}\) and \(\mathbf{A}_{agg} \in \mathbb{R}^{N \times C \times (K \times H \times  W)}\) represents the aggregated features of the support set.

{\bf Third Selection Step: Selective Map.} In order to select the key information for the target features, we compute the selective map based on \(\mathbf{F}_x\) and \(\mathbf{F}_{\text{sup}}\). First, \(\mathbf{F}_x\) and \(\mathbf{F}_{\text{sup}}\) are flattened into \(\mathbf{\hat{F}}_x \in \mathbb{R}^{ C \times (H \times W)}\) and \(\mathbf{\hat{F}}_{\text{sup}} \in \mathbb{R}^{ C \times (K  \times H \times W)}\), respectively. Then, the map is calculated through matrix multiplication, following ~\cite{cheng2021rethinking} and using the following equation:
\begin{equation}
\mathbf{Selective} = \frac{2 \cdot (\mathbf{\hat{F}}_{\text{sup}}^T \cdot \mathbf{\hat{F}}_x) - \mathbf{\hat{F}}_{\text{sup}}^2}{\sqrt{C}} 
\end{equation}
where \(C\) is the channel dimension of \(\mathbf{\hat{F}}_{\text{sup}}\) and \( \mathbf{Selective}  \in \mathbb{R}^{ (K  \times H \times W) \times (H \times W)}\). Afterward, the computed \(\mathbf{Selective}\) is normalized using the softmax to ensure that the contribution values conform to a probability distribution. The resultant information matrix can be represented as follows:
\begin{equation}
\mathbf{E}_{\text{ctx}} = \mathbf{A}_{agg} \cdot Softmax(\mathbf{Selective})
\end{equation}
where \(\mathbf{E}_{\text{ctx}}\) serves as the contextual (ctx) embedding of the key information from the support set, and \(\mathbf{E}_{\text{ctx}} \in \mathbb{R}^{N \times C \times (H \times W)}\).

\subsection{Contextual Colorization Module}
\label{sec:Colorization}
The CCM is proposed to interpret user intent from the contextual embedding derived from the support set, enabling human-model interaction. Conceptually, this process is similar to ``colorizing'' the target image based on the support mask. Unlike the CSM, which focuses on selecting contextually representative information tailored to the target image, this module emphasizes dynamically refining target features on the ground of the contextual embedding via cross-attention, thereby facilitating a deeper understanding of the user’s segmentation intent. Before entering the module, \(\mathbf{\hat{F}}_{\text{x}}\) is duplicated \(N\) times to align its dimensions with \(\mathbf{E}_{\text{ctx}}\).

The CCM consists of four identical blocks, one of which is detailed in \cref{fig:Figure3}. In each block, \(\mathbf{\hat{F}}_{\text{x}}\) and \(\mathbf{E}_{\text{ctx}}\) are each passed through a learnable projection layer to reduce their dimensions to \(\mathbf{\hat{F}}_{\text{x}} \in \mathbb{R}^{N \times C \times ((H \times W) / 2)}\) and \(\mathbf{E}_{\text{ctx}} \in \mathbb{R}^{N \times C \times ((H \times W) / 2)}\), respectively. Subsequently, \(\mathbf{E}_{\text{ctx}}\) is integrated into the target image features to guide the model in identifying specific regions expected to be ``colored''. This integration occurs by adding cross-attention-processed information back into \(\mathbf{\hat{F}}_{\text{x}}\), followed by a feed-forward layer (FFN)~\cite{vaswani2017attention} for contextual reasoning, yielding \(\mathbf{\hat{F}}_{\text{x\_next}}\), which then serves as the \textbf{next} target feature input for the next block. This process is represented as follows:
\begin{equation}
\mathbf{\hat{F}^{'}}_{\text{x}} = \text{Add\&Norm}(\text{Cross Attention}(\mathbf{\hat{F}}_{\text{x}}, \mathbf{E}_{\text{ctx}}))
\end{equation}
\begin{equation}
\mathbf{\hat{F}}_{\text{x\_next}} = \text{Add\&Norm}(\text{FFN}(\mathbf{\hat{F}^{'}}_{\text{x}}))
\end{equation}
Subsequently, \(\mathbf{E}_{\text{ctx}}\) reads from the updated target image features \(\mathbf{\hat{F}}_{\text{x\_next}}\) through a reversed cross-attention layer to pinpoint features essential for matching object information in the contextual embedding. In the followed FFN, the context embedding, equipped with object-specific feature representation, undergoes further enhancement to strengthen its segmentation guidance capability. The resulting \(\mathbf{E}_{\text{ctx\_next}}\) then serves as the \textbf{next} context embedding for the next block. This process can be expressed as follows:
\begin{equation}
\mathbf{E}^{'}_{\text{ctx}} = \text{Add\&Norm}(\text{Cross Attention}(\mathbf{E}_{\text{ctx}}, \mathbf{\hat{F}}_{\text{x\_next}}))
\end{equation}
\begin{equation}
\mathbf{E}_{\text{ctx\_next}} = \text{Add\&Norm}(\text{FFN}(\mathbf{E}^{'}_{\text{ctx}}))
\end{equation}
After passing through the final block, the iteratively refined context embedding fully comprehends the user-defined segmentation intent from the support set, as well as the corresponding image features derived from the target image. This enhanced context embedding subsequently serves as \(N\) prompts \(\mathbf{P}\) for the SAM or SAM 2 model, where \(\mathbf{P} \in \mathbb{R}^{N \times ((H \times W) / 16) \times C}\).

\subsection{Loss Function}
\label{sec:SAM 2andSAM}
For SAM, the image features \(\mathbf{F}_x\) extracted from the target image \(\mathbf{X}\) is fed into the Decoder with the prompt \(\mathbf{P}\) to generate the final output, represented by the following equation:
\begin{equation}
\text{Output} = \text{Decoder}(\mathbf{F}_x, \mathbf{P})
\end{equation}
where \(\text{Output} \in \mathbb{R}^{N \times H^0 \times W^0}\). For SAM 2, our approach directly inputs the high-dimensional vector \(\mathbf{P}\) into memory attention as prompt embeddings, thus \(\mathbf{P}\) also can be considered as memory extracted across video frames. For loss settings, we simply follow the design of SAM ~\cite{kirillov2023segment} to supervise the mask prediction with a Dice loss ~\cite{milletari2016v}, using the following formula:
\begin{equation}
\mathcal{L} = 1 - \frac{2 \sum_{i=1}^{D} p_i g_i}{\sum_{i=1}^{D} p_i + \sum_{i=1}^{D} g_i}
\end{equation}
where \(p_i\) and \(g_i\) represent the probability of pixel \(i\) in the predicted mask and the label of pixel \(i\) in the ground truth mask, respectively. The number of pixels involved in the computation of the Dice loss is denoted by \(D\).

\definecolor{gray}{rgb}{0.5, 0.5, 0.5} 
\begin{table*}[t]
\setlength{\abovecaptionskip}{1pt}
  \centering
  \begin{tabularx}{2.1\columnwidth}{@{}p{3cm}ccccccc@{}}
    \toprule
    \multirow{2}{*}{Method} & \multirow{2}{*}{\parbox{1cm}{\raggedright Training \\ Images}} & \multicolumn{3}{c}{Color Fundus Photography} & \multicolumn{2}{c}{Ultrasound} & \multirow{2}{*}{\parbox{2cm}{\raggedright Avg. Dice $\uparrow$ (\%)}} \\
    \cmidrule(lr){3-5} \cmidrule(l){6-7}
                            & & REFUGE2-Disc & REFUGE2-Cup & STARE-Vessel & FPA-PS & FPA-FH \\
    \midrule
    $Upper^{*}$ & $2~\text{x}~10^{3}$ & 93.7 & 83.5 & 65.5 & 82.1 & 91.7 & 83.3\\
    \midrule
    SAM$^{\text{TF}}$ (1-point) & - & 39.1 & 33.5 & 16.3 & 18.2 & 34.4 & 28.3\\
    SAM$^{\text{TF}}$ (box) & - & 54.2 & 71.6 & 20.5 & 67.0 & \underline{88.0} & 60.3\\
    SAM$^{\text{TF}}$ (everything) & - & 48.8 & 40.0 & 20.5 & 25.6 & 35.4 & 34.1\\
    MedSAM$^{\text{TF}}$ (box) & - & 87.2 & 81.3 & 20.1 & \textcolor{gray}{\bf 97.1} & \textcolor{gray}{\bf 97.5} & 74.5\\
    \midrule
    Med-SA (1-point) & 16 & {85.6} & {83.0} & {45.4} & {71.3} & {81.4} &{73.3}\\
    Med-SA (box) & 16 & {86.8} & {82.6} & {37.1} & {71.8} & {83.1} & {72.3}\\
    SAMed & 16 & {86.9} & {84.3} & {15.8} & {66.3} & 87.0 & {68.1}\\
    AutoSAM & 16 & \underline{87.9} & \underline{84.6} & \underline{73.8} & \underline{78.0} & 79.1 & \underline{80.7}\\
    \midrule
    Ours & 16 & {\bf 88.0} & {\bf 85.5} & {\bf 81.6} & {\bf 85.3} & {\bf 89.3} & {\bf 85.9}\\
    \bottomrule
  \end{tabularx}
  \caption{Comparison of our model with SAM-based SOTAs in Dice Score (\%) on image datasets, including train-free models (denoted as $^{\text{TF}}$), efficient fine-tuned models, and models trained on full data (denoted as $Upper^{*}$). ‘Training Images’ denotes the average number of images used for training to segment five objects, with dash(–) denoting inapplicability. \textcolor{gray}{Gray} indicates the data were used in model pre-training. The $\bf best$ and $\underline{\text{second-best}}$ results (excluding $Upper$) are bolded and underlined, respectively, showcasing our SOTA performance among various models under same few-shot settings.}
  \label{tab:compare_image}
  \vspace{-20pt}
\end{table*}
\definecolor{gray}{rgb}{0.5, 0.5, 0.5}
\begin{table}[t]
\setlength{\abovecaptionskip}{1pt}
\captionsetup{font=footnotesize}
  \centering
  \begin{tabularx}{1\columnwidth}{@{}p{2cm}ccc@{}}
    \toprule
    \multirow{2}{*}{Method} & \multicolumn{2}{c}{Training data} & \multirow{2}{*}{\parbox{1.5cm}{\raggedright Avg. Dice $\uparrow$ (\%)}}\\
    \cmidrule(lr){2-3}
     & With Label & W/O. Label &  \\
    \midrule
    $Upper^{*}$ & 1134 (100\%) & 0 (0\%) & 88.1\\
    \midrule
    DTC & 58 (5.1\%) & 1076 (94.9\%) & 63.4\\
    MLB-Seg & 58 (5.1\%) & 1076 (94.9\%) & 78.3\\
    \midrule
    nnUnet & 58 (5.1\%) & 0 (0\%) & 84.1\\
    Med-SA  & 58 (5.1\%) & 0 (0\%) & 76.0\\
    SAMed & 58 (5.1\%) & 0 (0\%) & {85.9}\\
    AutoSAM & 58 (5.1\%) & 0 (0\%) & 70.1\\
    \midrule
    Ours & 58 (5.1\%) & 0 (0\%) & {\bf 87.4}\\
    \bottomrule
  \end{tabularx}
  \caption{
 Comparison of our model against various SOTAs on MRI dataset, including semi-supervised models~\cite{luo2021semi, wei2023consistency}, SAM-based models, traditional segmentation models, and 3D DSD-FCN~\cite{wang2019deeply} trained on the full dataset (denoted as $Upper^{*}$). ‘Training Data’ indicates the number of slices used for training and their percentage of the total dataset.
  }
  \label{tab:compare_image3D}
  \vspace{-10pt}
\end{table}
\section{Experiment}
\label{sec:experiment}
\subsection{Datasets and Implementation Details}
We conducted extensive experiments on five widely used publicly available datasets, including REFUGE2~\cite{fang2022refuge2}, STARE~\cite{hoover2000locating}, FH-PS-AOP(FPA)~\cite{jieyun_2023_7851339}, PROMISE12~\cite{litjens2014evaluation} (3D MRI) and JNU-IFM~\cite{LU2022107904} (Video). Among these, STARE~\cite{hoover2000locating} and PROMISE12~\cite{litjens2014evaluation} contain only one single object, while the other datasets contain multiple segmentation objects. Further introduction in the supplementary.

\begin{table}[t]
\centering
\setlength{\abovecaptionskip}{1pt}
\begin{tabularx}{1\columnwidth}{p{1.5cm}XXXXX}
\toprule
Method & Prompt & {\raggedright Preload\\Capable} &  {\raggedright Avg.$\uparrow$\\(\%)} & {\raggedright Std.$\downarrow$\\(\%)} & {\small Max-Min}$\downarrow$(\%) \\ \hline
 \textcolor{gray}{$Upper$} & \textcolor{gray}{Mask} & \centering\textcolor{gray}{\ding{55}} & \textcolor{gray}{89.6} & - & - \\ 
\midrule
\rowcolor{green!5} SAM2 & {Box} & \centering\ding{55} & \underline{78.6} & - & - \\ 
\rowcolor{green!15} SAM2 & {Point} & \centering\ding{55} & 25.1 & - & - \\ 
\midrule
\rowcolor{green!25} MedSAM2 & {Mask} & \centering\ding{51} & 63.3 & \underline{10.9} & \underline{29.1} \\ 
\rowcolor{green!25}Ours & {Mask} & \centering\ding{51} & {\bf 80.9} & {\bf 0.3} & {\bf 1.0} \\ 
\bottomrule
\end{tabularx}
\caption{Comparison of our model with SAM2-based SOTAs on the video dataset, reporting Average Dice, standard deviation Dice, and max-min Dice difference. $Preload~Capable$ indicates whether the prompt can be pre-prepared for real-time use. Background shading from \colorbox{green!10}{light} to \colorbox{green!20}{dark green} represents increasing real-time usability, while SAM2-Mask, being nearly impractical, is treated as $Upper$ and marked in \textcolor{gray}{Gray}. The $\bf best$ and $\underline{\text{second-best}}$ results (excluding $Upper$) are bolded and underlined, respectively.
}
\label{tab:compare_Video}%
\vspace{-20pt}
\end{table}
Due to the limited training data, we freeze the pre-trained parameters of SAM to avoid overfitting. We applied LoRA~\cite{hu2021lora} to the original encoder and decoder for efficient fine-tuning on medical datasets. Model optimization was performed using the Stochastic Gradient Descent (SGD) optimizer, with a momentum of 0.9, a learning rate of 0.01, and a weight decay of 0.0005. For image and video datasets, we randomly selected 16 patient-level images or videos for training, with the remaining samples used for testing. For 3D data, 58 labeled slices from 3 patients were used for training, while data from 10 additional patients were reserved for testing. Further details in the supplementary.
\begin{figure*}
  \centering
  \setlength{\abovecaptionskip}{1pt}
   \includegraphics[width=0.9\linewidth]{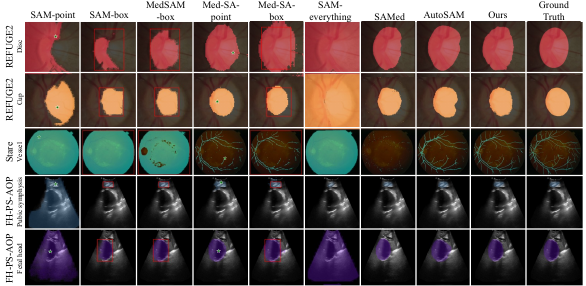}
   \caption{Visualization comparison results of nine models across five objects.
   }
   \label{fig:compare_Image}
   \vspace{-20pt}
\end{figure*}
\subsection{Comparison with SOTA on Image Dataset}
We extensively evaluated diverse comparison methods, including train-free foundation models, models efficiently fine-tuned on the specific-dataset, and traditional models ($Upper$). Details are in the supplementary. MedSAM, pretrained on a large medical dataset, was evaluated on both its seen FPA dataset and the unseen REFUGE2 dataset. For trainable models (Med-SA, SAMed, AutoSAM, and ours), we trained each on the same 16 images until convergence.

{\bf As shown in \cref{tab:compare_image}}, MedSAM and AutoSAM showed strong performance in both train-free and few-shot settings, probably due to pre-training on large medical datasets or object-specific training. However, MedSAM's performance dropped significantly on unseen datasets and struggled with intricate structure such as vessel in the STARE dataset. AutoSAM, on the other hand, underperformed on ultrasound datasets, possibly due to the challenges such as ultrasound imaging-associated speckle noise. Our method achieved SOTA performance under limited conditions, even comparable with traditional SOTAs trained on the full dataset. \cref{fig:compare_Image} shows that while SAMed and AutoSAM performed relatively well on REFUGE2 and FPA, both still struggled with boundary omissions. On STARE, MedSAM's box prompt led to segmentation ambiguity. Across all datasets, our method delivered consistently superior and stable results.
\subsection{Comparison with SOTA on MRI Dataset}
Although not specifically designed for 3D data, our model shows strong potential. We evaluated it on the 3D MRI PROMISE12 dataset~\cite{litjens2014evaluation} (details in the supplementary). {\bf As shown in \cref{tab:compare_image3D}}, it achieved the highest few-shot performance (87.4\%), surpassing all SAM-based models and the traditional nnUnet under the same conditions. Notably, despite not leveraging any additional unlabeled data, our method surpassed the best semi-supervised model (MLB-Seg~\cite{wei2023consistency}, 78.3\%). Furthermore, we achieved results comparable to the fully-supervised $Upper^{*}$ (88.1\%), with only a 0.7\% gap, showcasing its promise for 3D segmentation.

\subsection{Comparison with SOTA on Video Dataset}

We conducted a comprehensive comparison on the JNU-IFM dataset across four models: MedSAM2, SAM2-box, SAM2-point, and Ours. Since Ours and MedSAM2 allow pre-prepared prompts from non-target data, they offer the simplest real-time usability, followed by SAM2-point (one-click), and lastly SAM2-box. Although mask prompts provide the most detailed information, they are impractical for a single operator in real-time and thus serve as our $Upper$. To assess the stability of MedSAM2 and Ours, we conducted five trials on the same test set, each using a different support pair randomly drawn from five pre-excluded videos. Further experimental details are in supplementary.
{\bf As shown in \cref{tab:compare_Video}}, our method achieves the highest average Dice, improving by 17.6\% over MedSAM2, which uses a simple prompt strategy, and by 2.3\% over SAM2-box, which requires the most demanding prompt strategy. Further, our method yields comparable performance to the $Upper$, and presents excellent stability indicated by 0.3\% Std. Dice (far greater than MedSAM2 of 10.9\% Std. Dice and 29.1\% fluctuation in Dice). 



\begin{table}[t]
\footnotesize
    \centering
    \setlength{\abovecaptionskip}{1pt}
  \begin{tabular}{@{}cc|cc@{}}
    \toprule
    CSM & CCM & Avg. Dice $\uparrow$ (\%) & std. Dice $\downarrow$ (\%)\\
    \hline
    \ding{51} & \ding{51} & {\bf 85.9} & {\bf 2.6} \\
    \ding{51} & \ding{55} & 85.2 & 3.1\\
    \ding{55} & \ding{51} & 84.3 & 3.0\\
    \ding{55} & \ding{55} & 84.0 & 3.1\\
    \bottomrule
  \end{tabular}
  \caption{Ablation experiments on different PPG modules.}
  \label{tab:ablation_component}
  \vspace{-10pt}
\end{table}
\begin{figure}[t]
  \centering
  \setlength{\abovecaptionskip}{1pt}
   \includegraphics[width=0.99\columnwidth]{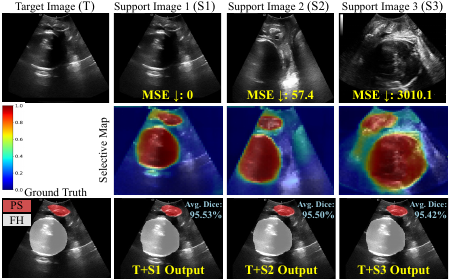}
   \caption{Selective Map visualization on support images with different mean square errors (MSE) to the target image.
   }
   \label{fig:visualization}
   \vspace{-10pt}
\end{figure}

\subsection{Ablation Studies}
We analyze the impact on PP-SAM from multiple aspects, including model architecture, pretrained weights, prompt quality and quantity during inference, and training data size. Experiment settings and details are in the supplementary.

{\bf Effectiveness of each model component.} 
The results in \cref{tab:ablation_component} highlight the contributions of both CSM and CCM to the optimal model performance. When CCM was removed, the segmentation performance std increased significantly (2.6\% → 3.1\%), indicating inconsistencies in multi-object segmentation. Removing CSM led to an overall degradation in segmentation quality, with the average Dice score decreasing by 1.6\%. When CSM and CCM were simultaneously omitted, the model achieved the lowest Avg. Dice (84.0\%) with the highest std. Dice (3.1\%), confirming that CSM and CCM collectively enhance both segmentation accuracy and stability. 

{\bf Visualization of selective map effectiveness across different support images.} 
\cref{fig:visualization} shows that the Selective Map directs the target image to focus on relevant anatomical structures in the support image while suppressing background interference. Across varying support image similarity, from S1 (MSE = 0) to S2 (MSE = 3010.1), it consistently highlights semantically relevant regions, ensuring stable segmentation (Avg. Dice: 95.42\%–95.53\%).

{\bf Effect of encoder architectures in CSM shown in \cref{tab:ablation_backbone}.} Vision Transformer achieves better results than ResNet-50 (Avg. Dice: 84.5\% vs. 83.9\%), which may be attributed to significantly larger number of trainable parameters (89.0M vs. 8.5M). Notably, Vision Mamba attains the highest performance (Dice: 85.9\%; IoU: 76.3\%) despite having the lowest parameter count (7.9M).
\begin{table}[t]
\setlength{\abovecaptionskip}{1pt}
\footnotesize
  \centering
  \begin{tabularx}{1\columnwidth}{@{}p{2.7cm}XXX@{}}
    \toprule
    Encoder in CSM & Turnable Param (M) & Avg. Dice $\uparrow$ (\%) & Avg. IoU $\uparrow$ (\%) \\
    \midrule
    Vision Mamba & {\bf 7.9} & {\bf 85.9} & {\bf 76.3} \\
    Vision Transformer & 89.0 & {84.5} & {74.0} \\
    ResNet-50 & 8.5 & {83.9} & {72.6} \\
    \bottomrule
  \end{tabularx}
  \caption{Ablation on different encoder architectures in CSM.}
  \label{tab:ablation_backbone}
  \vspace{-20pt}
\end{table}

{\bf Ablation on pretrained foundation models.} \cref{fig:ablation_pretrain} shows that our Proxy Prompt consistently enhances SAM-based models, including MedSAM, which was pretrained on extensive medical data. Surprisingly, MedSAM with our prompt underperforms compared to SAM with the same prompt, likely due to its strong reliance on Box prompts during pretraining, limiting the adaptability to our prompt. Deep analysis is in the supplementary.

\begin{table}[t]
\setlength{\abovecaptionskip}{1pt}
\footnotesize
  \centering
  \begin{tabularx}{1\columnwidth}{p{1cm}|p{1.9cm}|X|X}
    \toprule
    Method & Pretrained Weight & Prompt Method & Avg. Dice $\uparrow$ (\%) \\
    \hline
    \multirow{2}{*}{MedSAM} & \multirow{2}{*}{MedSAM-ViT-B} & Box & 74.5 \\
     &  & \cellcolor{gray!10}Proxy Prompt & \cellcolor{gray!10}81.3 \textcolor{green}{(+6.8)} \\
    \hline
    \multirow{4}{*}{SAM}& \multirow{2}{*}{SAM-ViT-B} & Box & {56.1} \\
    &  & \cellcolor{gray!10}Proxy Prompt & \cellcolor{gray!10}85.1 \textcolor{green}{(+29.0)} \\
    \cline{2-4}
    & \multirow{2}{*}{SAM-ViT-H} & Box & 60.3 \\
    &  & \cellcolor{gray!10}Proxy Prompt & \cellcolor{gray!10}{\bf 85.9} \textcolor{green}{(+25.6)} \\
    \bottomrule
  \end{tabularx}
  \caption{Performance with different pretrained foundation models.}
  \label{fig:ablation_pretrain}
  \vspace{-10pt}
\end{table}

{\bf Influence of prompt quality via support mask variations.} We evaluate six levels of support mask quality, recording each mask’s Dice score as Prompt Dice in \cref{fig:Diffmask}. The results shows segmentation improves (visualized by intensifying red shades) with prompt quality, dropping sharply only in $Empty$ level. Our model remains robust regardless of whether masks come from normal users or senior doctors. Line chart is in the supplementary.
\begin{table}[t]
\setlength{\abovecaptionskip}{1pt}
\footnotesize
  \centering
  \begin{tabularx}{1\columnwidth}{@{}p{2.3cm}XXXXXX@{}}
    \toprule
     & Empty & Low & Medium & High & SD & NU \\
    \midrule
    Prompt Dice (\%)& 00.0 & 66.6 & 81.6 & 100 & 88.3 & 74.8\\
    Output Dice (\%)& \colorbox{red!10}{47.9} & \colorbox{red!20}{77.8} & \colorbox{red!30}{86.5} & \colorbox{red!40}{86.8} & \colorbox{red!30}{86.6} & \colorbox{red!30}{86.5}\\
    \bottomrule
  \end{tabularx}
  \caption{Model performance across prompt qualities, including senior doctor (SD) and normal user (NU). Prompt Dice reflects support mask quality, while Output Dice measures performance.}
  \label{fig:Diffmask}
  \vspace{-10pt}
\end{table}
\begin{figure}[t]
  \centering
  \setlength{\abovecaptionskip}{1pt}
   \includegraphics[width=0.9\columnwidth]{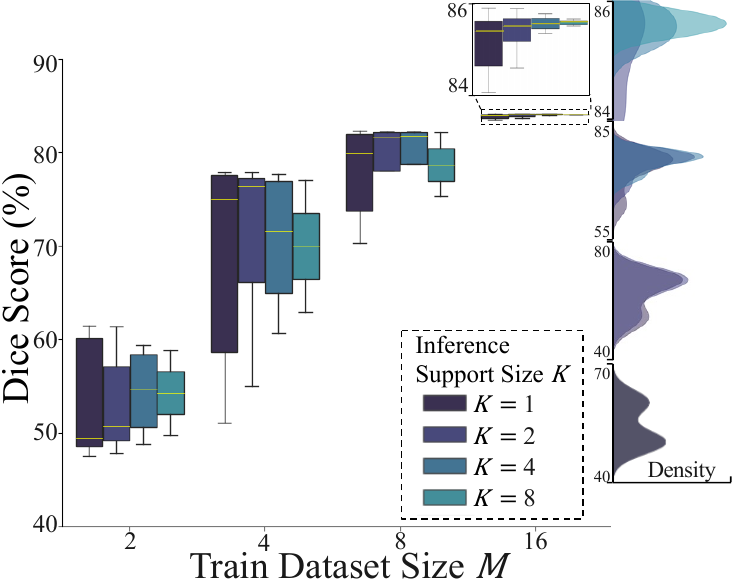}
   \caption{The boxes show performance for each \(M\) with varying \(K\), while in the rightmost plots (\(G < M-1\)), higher peaks refer to better performance and narrower ranges reflect improved stability. Optimum is achieved when both \(M\) and \(K\) are maximized.}
   \label{fig:ablation}
   \vspace{-20pt}
\end{figure}

{\bf Impact of training data size \( M \) and support pair quantity \( K \) in inference.} \cref{fig:ablation} shows that performance improves with \( M \). For each \( M \), Dice increases with \( K \) until \( K \) exceeds \( M-1 \), after which a slight decline may occur. The \( M=16 \) ridge plot indicates that as \( K \) increases, prediction variance decreases significantly. Overall, both training dataset size \( M \) and support size \( K \) influence model performance, which stabilizes at its peak when both are maximized. Experiment details are in the supplementary.

\section{Conclusion}
\label{sec:conclusion}
We present the PPG, a pluggable framework designed to endow SAM and SAM 2 with auto-prompting and enhanced their interactive capabilities. Our PPG employs a CSM to extract contextual information from non-target data and a CCM to interpret user intentions, enabling it to generate efficient PP that effectively guides the model output to meet user needs. 
{
   \small
   \bibliographystyle{ieeenat_fullname}
   \bibliography{main}
}

\clearpage
\begin{figure*}[t]
  \centering
  \setlength{\abovecaptionskip}{0pt}
   \includegraphics[width=0.99\linewidth]{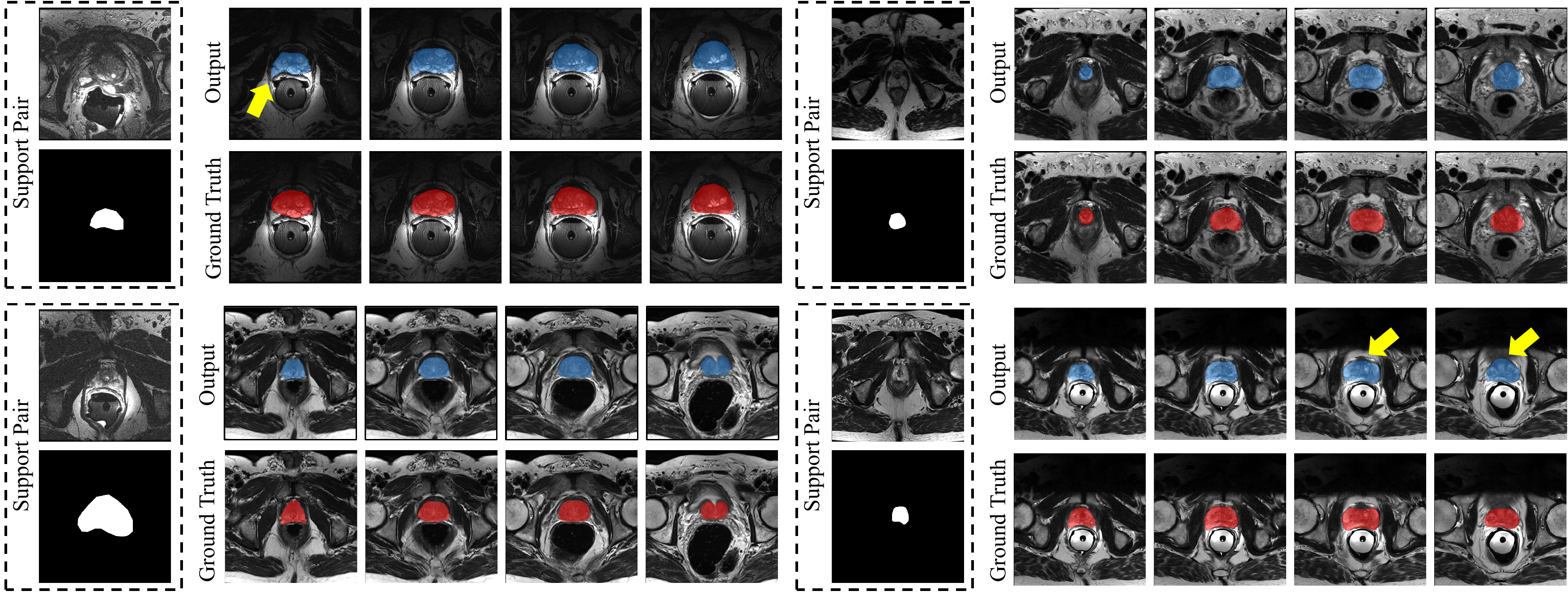}
   \caption{Quantitative comparison results on four representative examples.}
   \label{fig:compare_MRI}
   \vspace{-15pt}
\end{figure*}
\section{Prompt Strategy}
\label{sec:prompt strategy}
\subsection{Settings for Image Dataset.} 
For the models that require prompts in the comparison experiments on image dataset, the following prompt conditions are provided.

1. {\bf Point}: Since the center point of the disc and vessel mask is not on the target, one random point within the target mask as a positive point prompt. 

2. {\bf Box}: Minimum bounding rectangle of the target as a bounding box prompt. 

3. {\bf Everything}: Automatically segment multiple targets with everything mode. We select the prediction that the highest overlaps with ground truth to calculate the model's Dice score. 

4. {\bf Others}: The SAMed and AutoSAM models are designed to perform automatic segmentation without manual-given prompt, while we use the support image-mask pair as a prompt.

\subsection{Settings for Video Dataset.}
Considering the high demand for timely operation in real-time image-guided examinations and interventions for physicians, prompt was only required for the first frame of the video data. 

1. {\bf Box}: Minimum bounding rectangle of the target as a bounding box prompt. 

2. {\bf Point}: Center point of the object as a positive point. 

3. {\bf Mask}: Target mask of the patient under examination.

4. {\bf Support Image-Mask Pair}: Image of the first frame and the corresponding target mask from ``non-target'' data (i.e., image/video frame of subjects other than the one under examination, such as from retrospective datasets).

\begin{table}[t] 
  \centering
  \setlength{\abovecaptionskip}{1pt}
  \begin{tabular}{@{}p{1.3cm}p{1.5cm}p{2.4cm}p{1.9cm}@{}} 
    \toprule
    Dataset & Modality & Segmentation Objects & Samples \\
    \midrule
    \rowcolor{orange!10} \raggedright REFUGE2~\cite{fang2022refuge2} & \raggedright Fundus & \raggedright Optic disc and optic cup & 2000 images \\
    \rowcolor{teal!30} \raggedright STARE~\cite{hoover2000locating} & \raggedright Fundus & \raggedright Blood vessels in retinal images & 20 images \\
    \rowcolor{blue!10} \raggedright FPA~\cite{jieyun_2023_7851339} & \raggedright Ultrasound & \raggedright Fetal head and pubic symphysis & 4000 images \\
    \rowcolor{cyan!10} \raggedright PROMISE12~\cite{litjens2014evaluation} & \raggedright MR Images & \raggedright Prostate & \parbox[t]{1.9cm}{50 3D transversal T2-weighted MRI} \\
    \rowcolor{yellow!10} \raggedright JNU-IFM~\cite{LU2022107904} & \raggedright Ultrasound & \raggedright Fetal head and pubic symphysis & 78 videos \\
    \bottomrule
  \end{tabular}
  \label{tab:dataset}
  \caption{Dataset Summary.}
  \vspace{-20pt}
\end{table}
\section{Datasets Introduction}
The type of image modalities, segmentation objects, and number of samples for the four included datasets are summarized in the table below. REFUGE2~\cite{fang2022refuge2}, STARE~\cite{hoover2000locating} and FPA~\cite{jieyun_2023_7851339} are image datasets evaluated in the image segmentation task, whereas PROMISE12~\cite{litjens2014evaluation} and JNU-IFM~\cite{LU2022107904} serve as 3D and video datasets, evaluated in MRI and video segmentation tasks, respectively.

\subsection{Implementation Details}
The SAM model was used with pre-trained weights, while the Vision Mamba encoder was trained from scratch without pre-trained parameters. The patch size and embedding dimension of the ViM~\cite{zhu2024vision} encoder were set to 16 and 192, respectively.  

For support pairs in inference, they are randomly selected from the training dataset. Image, 3D, and video tasks were trained separately. For the video dataset, we trained a single model capable of segmenting either the fetal head or pubic symphysis based on user instructions. For the 3D dataset, training was conducted on slices from three patients at varying depths, while testing was performed on slices from ten independent patients. For the image dataset, we trained two models based on modality: one for STARE and REFUGE2 and another for the remaining datasets.  

In fact, training our models separately for each dataset yields better performance. However, considering real-world clinical applications, providing a more user-friendly model is preferred for us when accuracy loss is not that much (0.5\%). This allows physicians to customize segmentation by simply adjusting the support mask (\eg disc or vessel) on the same image without retraining. Therefore, we trained models per modality to better simulate real-world usage across medical specialties. The impact of different training strategies on image datasets is presented in \cref{exp:Diff_train}.
During training, we alternate data within same dataset as either the support image or test image, enabling the model to learn how to extract prompts from the support pair. Additionally, for the same support image A, we vary the support mask to indicate different objects, training the model to segment multiple targets objects the same target image B.

\subsection{Experiment Details: Comparison with SOTA on Image Dataset}
\label{exp:Diff_train}
For train-free models such as SAM and MedSAM, we tested their performance with different prompt.  For traditional segmentation models, we used dataset-specific SOTA methods, including BEAL~\cite{wang2019boundary} for REFUGE2, nnUnet~\cite{isensee2021nnu} for STARE, and Segnet~\cite{badrinarayanan2017segnet} for FPA. The results of the three models, as claimed in their papers, were obtained under full data training settings and thus were treated as the $Upper$.
We evaluated our model’s performance under three training strategies: training each dataset independently (Strategy 1), training by modality (Strategy 2), and training on all datasets combined (Strategy 3). As shown in \cref{tab:Diff_train}, Strategy 1 achieves the best performance overall, likely because training on a single dataset avoids interference from others. Comparing Strategies 2 and 3, we observe an increase in Dice scores for ultrasound datasets under Strategy 3, but a sharp performance drop on STARE (81.6\% → 74.2\%), suggesting that joint training may compromise segmentation on certain datasets. While the results of Strategies 2 and 3 are comparable, Strategy 3 falls slightly behind in overall performance.  

From a clinical perspective, specialists often prefer a highly customized model that delivers precise segmentation for their specific area of interest with minimal training effort. For instance, ophthalmologists prioritize high accuracy in fundus imaging, while performance on ultrasound data is less relevant to their workflow. However, training a separate model for each segmentation target increases their operational burden, making this trade-off nontrivial. Balancing these factors, we report results under Strategy 2, which offers a practical compromise between accuracy and usability.

\definecolor{gray}{rgb}{0.5, 0.5, 0.5} 
\begin{table*}[t]
\setlength{\abovecaptionskip}{1pt}
  \centering
  \begin{tabularx}{2.1\columnwidth}{@{}p{2cm}ccccccc@{}}
    \toprule
    \multirow{2}{*}{Method} & \multirow{2}{*}{\parbox{1cm}{\raggedright Training \\ Strategy}} & \multicolumn{3}{c}{Color Fundus Photography} & \multicolumn{2}{c}{Ultrasound} & \multirow{2}{*}{\parbox{2cm}{\raggedright Avg. Dice $\uparrow$ (\%)}} \\
    \cmidrule(lr){3-5} \cmidrule(l){6-7}
                            & & REFUGE2-Disc & REFUGE2-Cup & STARE-Vessel & FPA-PS & FPA-FH \\
    \midrule
    \multirow{3}{*}{Ours} & 1 & {\bf 88.3} & {\bf 86.0} & {\bf 82.5} & {\bf 85.7} & {\bf 89.5} & {\bf 86.4}\\
     & 2 & \underline{88.0} & \underline{85.5} & \underline{81.6} & {85.3} & {89.3} & \underline{85.9}\\
     & 3 & {87.2} & {84.9} & {74.2} & \underline{85.6} & \underline{89.7} & {84.3}\\
    \bottomrule
  \end{tabularx}
  \caption{Model performance under different strategy.}
  \label{tab:Diff_train}
  \vspace{-20pt}
\end{table*}
\section{Experiment Details: Comparison with SOTA on MRI Dataset}
We extensively evaluated various methods, including semi-supervised models (DTC, MLB-Seg), SAM-based models (Med-SA, SAMed, AutoSAM), traditional segmentation models (nnUNet), and 3D DSD-FCN~\cite{wang2019deeply}, trained on the full dataset (denoted as $Upper^{*}$). 3D DSD-FCN~\cite{wang2019deeply} and MLB-Seg are existing SOTA models for PROMISE12~\cite{litjens2014evaluation} dataset under fully supervised and semi-supervised settings, respectively.
For semi-supervised training, 58 labeled slices from 3 cases and 1076 unlabeled slices from 37 cases were used for training until convergence. In contrast, nnUNet and all SAM-based models were trained on the same 58 labeled slices until convergence. All methods share identical data preprocessing, and evaluation was conducted on the same 10 test cases. Med-SA utilized the minimum bounding rectangle of the object as a bounding box prompt, while our method used a support image-mask pair as the prompt.

{\bf The qualitative comparison results are shown in \cref{fig:compare_MRI}.} These four representative examples are selected from multiple slices across four test cases, with each support slice randomly chosen from the training set along with its corresponding prostate annotation. Yellow arrows indicate minor segmentation overflow or omission. In the bottom-right example, the highlighted region shows that the model segments based solely on the current slice, disregarding consistency with adjacent slices. One possible reason is that the pretrained model has limited exposure to 3D data and, with only 58 slices, has primarily adapted to the MRI modality rather than developing a true 3D understanding. Another reason could be that the current architecture is not explicitly designed for 3D data, making it less capable of capturing spatial relationships across slices. Nevertheless, most prostate regions are precisely segmented, demonstrating the strong capability of our model and its potential for 3D data.

\section{Experiment Details: Comparison with SOTA on Video Dataset}
Beyond evaluating our model on image and 3D datasets, we further investigated its performance on video data. Given that SAM2 is pretrained on a large-scale video dataset, it was a natural choice to integrate our method into SAM2 for video segmentation. In this present work, we conducted a comprehensive comparison on the JNU-IFM dataset across four models: MedSAM 2, SAM 2-box, SAM 2-point, and Ours. 16 videos were randomly selected for training, with the remaining used for testing. The test set was identical for all models.

\subsection{Fairer Comparison: Mask Prompt} Models using mask prompts generally achieve superior performance as they inherently provide more information than point or box prompts. To ensure a fairer comparison, we also applied mask prompts to SAM2 for segmentation. However, requiring mask prompts to be manually provided on target data in real-time is impractical for a single operator, especially in ultrasound imaging, where the view dynamically shifts with probe movement. Thus, this setting serves as our $Upper$ bound. Further, we aimed to compare other methods capable of utilizing non-target image-mask prompts for SAM-based segmentation, testing their feasibility under practical conditions. To the best of our knowledge, there are currently no existing models leveraging non-target image-mask pairs to prompt SAM2 for video segmentation. We then observed that MedSAM2’s core idea treats 2D and 3D data as a video stream, using a template as an initial frame, which can originate from non-target data. Based on this, we employed our support pair as MedSAM2’s template and evaluated its segmentation performance. Since MedSAM2 follows a OnePrompt Segmentation approach, we ensured a fair comparison by restricting our method to a single support image-mask pair as the prompt.

\subsection{Model Stability Experiment Details} Since the support image-mask pair originates from non-target data, different support pairs may lead to varying model performance. We conducted five experiments on the same test set across the MedSAM2 and ours with five different support pairs to evaluate model performance and stability. We randomly excluded five videos from five patients and paired the first frame's image and mask as five candidate support pairs. In each experiment, we altered different support pair as prompt to assess stability.


{\bf The quantitative comparison results are shown in \cref{fig:bar_graph}.} The figure clearly shows that SAM2-Point underperforms significantly compared to other models, likely due to the insufficient information provided by point prompts in ultrasound images with blurred object boundaries. MedSAM2 demonstrates strong segmentation performance but exhibits high variability across different support pairs, sometimes falling below SAM2-Box while surpassing it in Support Pair 2. Further visualization of MedSAM2’s performance across Support Pair 1 and Support Pair 2 can be found in \cref{fig:compare_video}. SAM2-Box maintains stable Dice scores around 78\% but remains inferior to our model. {\bf Detailed per-object results can be found in \cref{tab:five_exp_ps} and \cref{tab:five_exp_fh}}.

{\bf The qualitative comparison results are shown in \cref{fig:compare_video}.} Support pair 1 and 2 were chosen to demonstrate the stability of MedSAM 2 and our model under different prompts. Due to the blurred FH boundary in this case, both SAM 2-box and Ours (as well as the $Upper$) exhibited slight over-segment in FH segmentation. However, while SAM 2-box over-segmented the FH region to the left in \(Frame_{t}\), our method constrains this expansion. Although MedSAM 2 showed a notable drop in PS segmentation performance with support pair 1, our method maintained consistent results

\afterpage{\clearpage
\begin{figure*}
  \centering
  \setlength{\abovecaptionskip}{3pt}
   \includegraphics[width=0.99\linewidth]{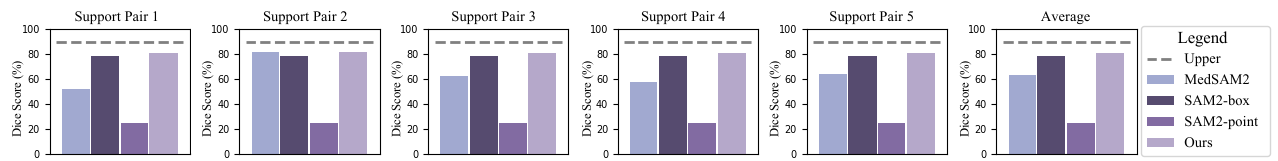}
   \caption{Comparison on the video dataset using five different support pairs, showing the average results for two objects. Results show that our method achieves SOTA performance over existing models and demonstrates stability with a maximum fluctuation of only 1.0\%.}
   \label{fig:bar_graph}
   \vspace{-15pt}
\end{figure*}
\begin{table*}
\centering
\setlength{\abovecaptionskip}{1pt}
\footnotesize
\begin{tabularx}{\textwidth}{p{1.2cm}cc|cc|cc|cc|cc|cc|c|c}
\toprule
\multirow{2}{*}{Method} & \multirow{2}{*}{Prompt} & \multirow{2}{*}{\parbox{1cm}{ Preload\\Capable}} & \multicolumn{2}{c}{S1} & \multicolumn{2}{c}{S2} & \multicolumn{2}{c}{S3} & \multicolumn{2}{c}{S4} & \multicolumn{2}{c}{S5} &  \multirow{2}{*}{\raggedright Avg. Dice$\uparrow$} & \multirow{2}{*}{\raggedright Avg. IoU$\uparrow$} \\
\cmidrule(lr){4-5} \cmidrule(lr){6-7} \cmidrule(lr){8-9} \cmidrule(lr){10-11} \cmidrule(lr){12-13}

& & & Dice & IoU & Dice & IoU & Dice & IoU & Dice & IoU & Dice & IoU \\
\hline
 \textcolor{gray}{SAM2} & \textcolor{gray}{Mask} & \centering\textcolor{gray}{\ding{55}} & \textcolor{gray}{87.9} & \textcolor{gray}{79.2} & \textcolor{gray}{87.9} & \textcolor{gray}{79.2} & \textcolor{gray}{87.9} & \textcolor{gray}{79.2} & \textcolor{gray}{87.9} & \textcolor{gray}{79.2} & \textcolor{gray}{87.9} & \textcolor{gray}{79.2} & \textcolor{gray}{87.9} & \textcolor{gray}{79.2}\\ 
\midrule
\rowcolor{green!5} SAM2 & {Box} & \centering\ding{55} & 76.8 & 64.2 & 76.8 & 64.2 & 76.8 & 64.2 & 76.8 & 64.2 & 76.8 & 64.2 & 76.8 & 64.2\\ 
\rowcolor{green!15} SAM2 & {Point} & \centering\ding{55} & 18.9 & 12.3 & 18.9 & 12.3 & 18.9 & 12.3 & 18.9 & 12.3 & 18.9 & 12.3 & 18.9 & 12.3\\ 
\midrule
\rowcolor{green!25} MedSAM2 & {Mask} & \centering\ding{51} & 50.8 & 37.2 & 77.9 & 64.8 & 67.4 & 54.3 & 60.9 & 46.4 & 56.2 & 43.1 & 62.6 & 49.2 \\ 
\rowcolor{green!25}Ours & {Mask} & \centering\ding{51} & {\bf 77.9} & {\bf 64.9} & {\bf 78.0} & {\bf 65.0} & {\bf 77.8} & {\bf 64.8} & {\bf 78.1} & {\bf 65.0} & {\bf 78.0} & {\bf 65.1} & {\bf 78.0} & {\bf 64.9} \\ 
\bottomrule
\end{tabularx}
\caption{
Comparison on the video dataset using five different support pairs, presenting Dice and IoU results for the pubic symphysis. $S1-S5$ denote the five support pairs, with Dice (\%) and IoU (\%) reported for each experiment and their averages. $Preload Capable$ indicates whether the prompt can be pre-prepared, allowing direct application in real-time surgeries. Since both MedSAM2 and Ours utilize non-target data as prompts, they enable preloading, offering greater convenience for real-time clinical use. To reflect ease of prompt availability in real-time scenarios, background shading from light to dark green represents increasing usability, while SAM2-Mask, being nearly impractical, is treated as $Upper$ and marked in gray.
}
\label{tab:five_exp_ps}%
\end{table*}
\begin{table*}
\centering
\setlength{\abovecaptionskip}{1pt}
\footnotesize
\begin{tabularx}{\textwidth}{p{1.2cm}cc|cc|cc|cc|cc|cc|c|c}
\toprule
\multirow{2}{*}{Method} & \multirow{2}{*}{Prompt} & \multirow{2}{*}{\parbox{1cm}{ Preload\\Capable}} & \multicolumn{2}{c}{S1} & \multicolumn{2}{c}{S2} & \multicolumn{2}{c}{S3} & \multicolumn{2}{c}{S4} & \multicolumn{2}{c}{S5} &  \multirow{2}{*}{\raggedright Avg. Dice$\uparrow$} & \multirow{2}{*}{\raggedright Avg. IoU$\uparrow$} \\
\cmidrule(lr){4-5} \cmidrule(lr){6-7} \cmidrule(lr){8-9} \cmidrule(lr){10-11} \cmidrule(lr){12-13}

& & & Dice & IoU & Dice & IoU & Dice & IoU & Dice & IoU & Dice & IoU \\
\hline
 \textcolor{gray}{SAM2} & \textcolor{gray}{Mask} & \centering\textcolor{gray}{\ding{55}} & \textcolor{gray}{91.3} & \textcolor{gray}{84.4} & \textcolor{gray}{91.3} & \textcolor{gray}{84.4} & \textcolor{gray}{91.3} & \textcolor{gray}{84.4} & \textcolor{gray}{91.3} & \textcolor{gray}{84.4} & \textcolor{gray}{91.3} & \textcolor{gray}{84.4} & \textcolor{gray}{91.3} & \textcolor{gray}{84.4} \\ 
\midrule
\rowcolor{green!5} SAM2 & {Box} & \centering\ding{55} & 80.3 & 71.1 & 80.3 & 71.1 & 80.3 & 71.1 & 80.3 & 71.1 & 80.3 & 71.1 & 80.3 & 71.1 \\ 
\rowcolor{green!15} SAM2 & {Point} & \centering\ding{55} & 31.2 & 18.6 & 31.2 & 18.6 & 31.2 & 18.6 & 31.2 & 18.6 & 31.2 & 18.6 & 31.2 & 18.6 \\ 
\midrule
\rowcolor{green!25} MedSAM2 & {Mask} & \centering\ding{51} & 52.9 & 42.0 & 84.2 & 73.4 & 56.6 & 42.9 & 54.8 & 44.3 & 71.5 & 59.0 & 64.0 & 52.3 \\ 
\rowcolor{green!25}Ours & {Mask} & \centering\ding{51} & {\bf 83.4} & {\bf 72.9} & {\bf 84.5} & {\bf 74.1} & {\bf 83.5} & {\bf 73.0} & {\bf 83.3} & {\bf 72.7} & {\bf 84.0} & {\bf 73.5} & {\bf 83.7} & {\bf 73.2} \\ 
\bottomrule
\end{tabularx}
\caption{
Comparison on the video dataset using five different support pairs, presenting Dice and IoU results for the fetal head. $S1-S5$ denote the five support pairs, with Dice (\%) and IoU (\%) reported for each experiment and their averages. $Preload Capable$ indicates whether the prompt can be pre-prepared, allowing direct application in real-time surgeries. Since both MedSAM2 and Ours utilize non-target data as prompts, they enable preloading, offering greater convenience for real-time clinical use. To reflect ease of prompt availability in real-time scenarios, background shading from light to dark green represents increasing usability, while SAM2-Mask, being nearly impractical, is treated as $Upper$ and marked in gray.
}
\label{tab:five_exp_fh}%
\end{table*}
\begin{figure*}
  \centering
  \setlength{\abovecaptionskip}{3pt}
   \includegraphics[width=0.99\linewidth]{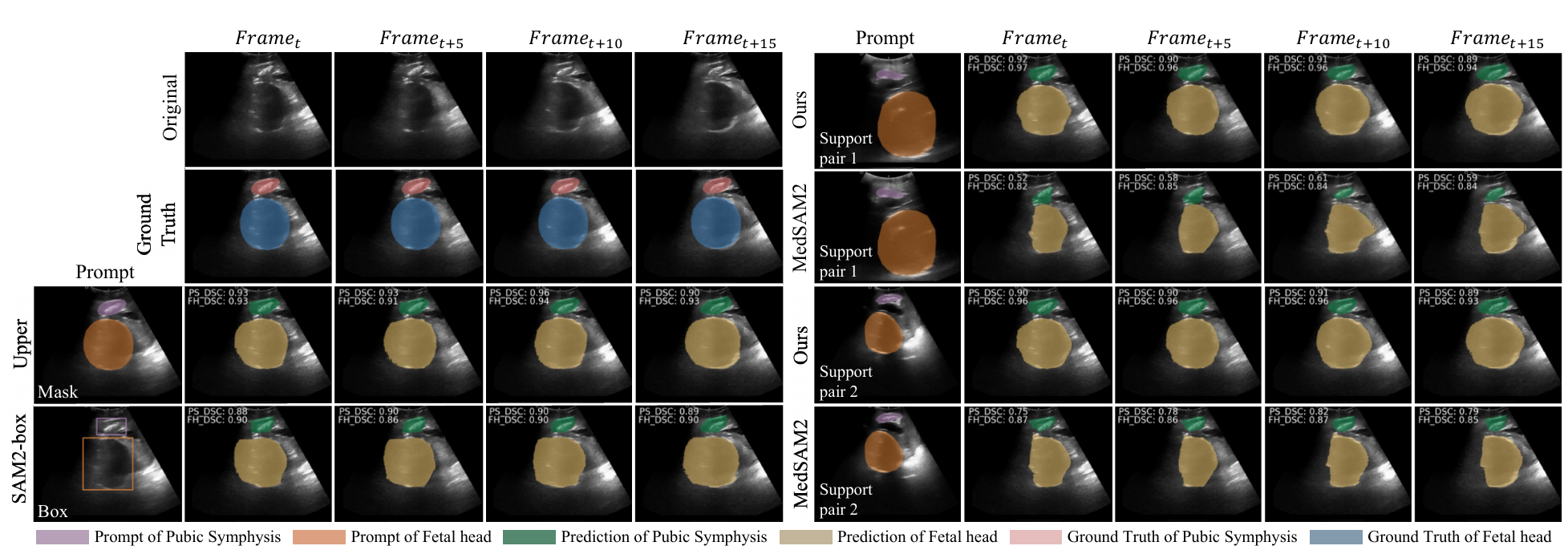}
   \caption{Comparison on one representative video. The top left corner of each subplot displays the Dice scores for the segmentation of the public symphysis(PS\_DSC) and fetal head (FH\_DSC).}
   \label{fig:compare_video}
   \vspace{-15pt}
\end{figure*}
\FloatBarrier
\clearpage
}
\section{Discussion of related work}
We also note that in the domain of natural image datasets, some works share a similar paradigm to ours~\cite{sun2024vrp}. However, our approach differs significantly from VRP-SAM~\cite{sun2024vrp} in both motivation and methodology. Specifically, our study addresses a clinical need where doctors must focus on different objects for different tasks. To accommodate this, we designed CCM, enabling segmentation of multiple objects within the same image using different prompts. In contrast, VRP-SAM does not explicitly tackle this challenge, likely because it is not a critical requirement in their problem setting.  

These differences stem from fundamentally distinct problem formulations. In clinical scenarios, large volumes of retrospective data exist, yet high-quality annotations are scarce. Meanwhile, doctors must frequently segment new incoming data, creating a significant workload. This motivated us to explore whether retrospective data, combined with limited labels, could help the model quickly generalize segmentation capabilities within similar modalities, thereby reducing the burden on clinicians. Given that both retrospective and newly generated data often belong to the same modality or anatomical structure, we leveraged Vision Mamba's input-invariant capability to adapt on modalities. This contrasts with natural image datasets, where prompts and target data are typically unrelated in structure but share similar foregrounds, making such design less necessary. These fundamental differences naturally led to divergent methodological designs.  

Beyond VRP-SAM~\cite{sun2024vrp}, various approaches use image-mask pairs as prompts, spanning train-free methods~\cite{zhang2023personalize, chen2024segmentation, xu2023eviprompt}, fine-tuning approaches~\cite{sun2024vrp}, non-SAM-based~\cite{yang2024tavp, li2024visual} and pretrained foundation models~\cite{zou2023segment, wu2024one, butoi2023universeg}. However, their similarities are limited to the paradigm itself, as each method is tailored to distinct objectives. To our knowledge, our non-target data (retrospective data) high-dimensional prompting strategy, as well as the CCM/CSM modules, which enable rapid adaptation to custom datasets with limited data, have not been explored in prior works.


\section{Detailed settings and results: Ablation study}
We structured our ablation studies into three key aspects: {\bf our proposed modules}, {\bf the retrospective image-mask pair prompt strategy}, and {\bf methodological parameters}.  

For {\bf our proposed modules}, we conducted ablations on {\bf CSM} and {\bf CCM}, visualized the effectiveness of the {\bf Selective Map}, and analyzed different {\bf encoder architectures} within {\bf CSM}.  

For {\bf our retrospective image-mask pair prompt strategy}, we examined the influence of {\bf support image relevance}, varying both its {\bf MSE with the target image} and testing support images from entirely unrelated datasets. We also evaluated six levels of {\bf prompt quality} by modifying {\bf support mask with different dice score} and analyzed the effect of different {\bf support pair quantities} on model performance.  

For {\bf methodological parameters},  \( N \) was fixed as the number of segmentation objects per task. We conducted ablations on {\bf training dataset size} (\( K \)) and {\bf support pair quantity during inference} (\( K \)). To evaluate \( K \), we tested it across four training dataset sizes (\( M = 2, 4, 8, 16 \)), setting the number of support pairs during training to \( M-1 \). Additionally, we ablated {\bf pretrained foundation models}, assessing how different initialization weights affected segmentation performance.
\subsection{Effectiveness of each model component.}
The results in \cref{tab:CSM_CCM_ablation} highlight the contributions of both CSM and CCM to the optimal model performance. When CCM was removed, the segmentation performance for PS and vessel structures showed a pronounced decline (1.6\% and 1.4\% Dice score drop). This may be attributed to the absence of CCM impairing the PPG’s ability to interpret different user intentions, resulting in markedly divergent segmentation outcomes under varying indications. On the other hand, removing CSM led to an overall degradation in segmentation quality for both structures, with the average Dice score decreasing by 1.6\%. When the selective mechanism of CSM was disabled and CCM was simultaneously omitted, the model's performance dropped to its lowest. This experiment demonstrates that our model, when enhanced with both modules, achieves optimal and consistent performance across different objects.

\begin{table*}[t]
\setlength{\abovecaptionskip}{1pt}
  \centering
  \begin{tabular}{@{}p{1cm}p{1cm}ccccccc@{}}
    \toprule
    \multirow{2}{*}{CSM} & \multirow{2}{*}{CCM} & \multicolumn{3}{c}{Color fundus photography} & \multicolumn{2}{c}{Ultrasound} & \multirow{2}{*}{Avg. Dice $\uparrow$ (\%)} \\
    \cmidrule(lr){3-5} \cmidrule(l){6-7}
                            & & REFUGE2-Disc & REFUGE2-Cup & STARE-Vessel & FPA-PS & FPA-FH \\
    \midrule
    \ding{51} & \ding{51} & {\bf 88.0} & {\bf 85.5} & {\bf 81.6} & {\bf 85.3} & {\bf 89.3} & {\bf 85.9}\\
    \ding{51} & \ding{55} & {87.6} & {85.3} & {80.2} & {83.7} & {89.2} & {85.2}\\
    \ding{55} & \ding{51} & 87.1 & 84.6 & 79.2 & {83.3} & {87.6} & {84.3}\\
    \ding{55} & \ding{55} & 87.0 & 83.7 & 79.0 & {82.7} & {87.6} & {84.0}\\
    \bottomrule
  \end{tabular}
  \caption{The detailed per-subdataset (subobject) results of the ablation experiments for both modules.}
  \label{tab:CSM_CCM_ablation}
  \vspace{-10pt}
\end{table*}

\begin{figure*}[t]
  \centering
  \setlength{\abovecaptionskip}{3pt}
   \includegraphics[width=0.99\linewidth]{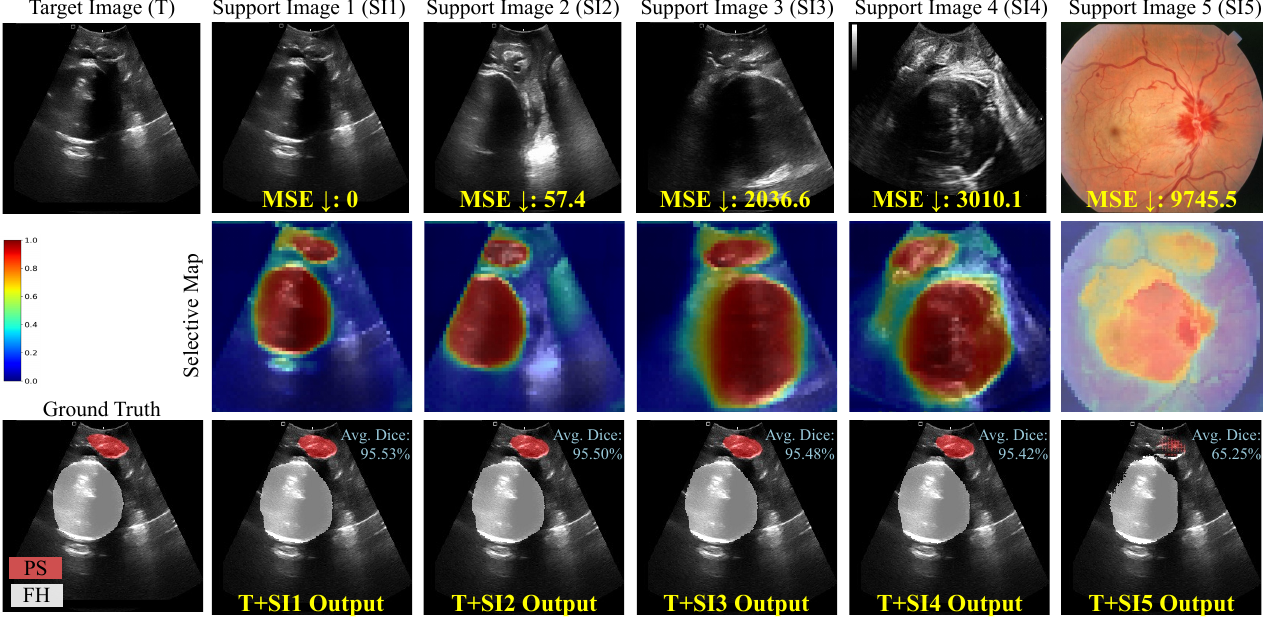}
   \caption{Visualization of Selective Map across different support images.}
   \label{fig:visual}
   \vspace{-15pt}
\end{figure*}
\begin{table*}[t]
\setlength{\abovecaptionskip}{1pt}
  \centering
  \begin{tabularx}{\textwidth}{@{}p{1cm}p{2.9cm}p{1.7cm}XXXXXp{1.7cm}@{}}
    \toprule
    \multirow{2}{*}{Method} & \multirow{2}{*}{Encoder in CSM} & \multirow{2}{*}{\parbox{1.7cm}{Turnable Param (M)}} & \multicolumn{3}{c}{Color fundus photography} & \multicolumn{2}{c}{Ultrasound} & \multirow{2}{*}{\parbox{1.7cm}{Avg. Dice $\uparrow$ (\%)}} \\
    \cmidrule(lr){4-6} \cmidrule(l){7-8}
                            & & & REFUGE2-Disc & REFUGE2-Cup & STARE-Vessel & FPA-PS & FPA-FH \\
    \midrule
    \multirow{3}{*}{Ours} & Vision Mamba & 7.9 & {\bf 88.0} & {\bf 85.5} & {\bf 81.6} & {\bf 85.3} & {\bf 89.3} & {\bf 85.9}\\
     & Vision Transformer & 89.0 & {86.8} & {84.3} & {79.6} & {83.2} & {88.4} & {84.5}\\
     & ResNet-50 & 8.5 & 85.8 & 83.9 & 79.3 & {82.4} & {88.0} & {83.9}\\
    \bottomrule
  \end{tabularx}
  \caption{The detailed per-subdataset (subobject) results of the ablation experiments for different Encoder.}
  \label{tab:ablation_backbone}
  \vspace{-10pt}
\end{table*}
\begin{table*}[t]
\setlength{\abovecaptionskip}{1pt}
\footnotesize
  \centering
  \begin{tabularx}{\textwidth}{@{}p{1cm}|p{2cm}|p{1cm}|X|X|X|X|X|X@{}}
    \toprule
    \multirow{2}{*}{Method} & \multirow{2}{*}{Pretrained Weight} & \multirow{2}{*}{\parbox{1cm}{Prompt Method}} & \multicolumn{3}{c|}{Color fundus photography} & \multicolumn{2}{c|}{Ultrasound} & \multirow{2}{*}{Avg. Dice $\uparrow$ (\%)} \\
    \cline{4-6} \cline{7-8}
                            & & & REFUGE2-Disc & REFUGE2-Cup & STARE-Vessel & FPA-PS & FPA-FH \\
    \hline
    \multirow{2}{*}{MedSAM} & \multirow{2}{*}{MedSAM-ViT-B} & Box & 87.2 & 81.3 & 20.1 & \textcolor{gray}{\bf 97.1} & \textcolor{gray}{\bf 97.5} & 74.5\\
    &  & \cellcolor{gray!10} PP & \cellcolor{gray!10}{\bf 94.3} \textcolor{green}{(+7.1)} & \cellcolor{gray!10}83.8 \textcolor{green}{(+2.5)} & \cellcolor{gray!10}59.4 \textcolor{green}{(+39.3)} & \cellcolor{gray!10}81.9 \textcolor{red}{(-15.2)} & \cellcolor{gray!10}86.9 \textcolor{red}{(-10.6)} & \cellcolor{gray!10}81.3 \textcolor{green}{(+6.8)} \\
    \hline
    \multirow{4}{*}{SAM}& \multirow{2}{*}{SAM-ViT-B} & Box & 39.3 & 68.2 & 20.3 & 69.0 & 83.9 & 56.1 \\
    &  & \cellcolor{gray!10}PP & \cellcolor{gray!10}87.5 \textcolor{green}{(+48.2)} & \cellcolor{gray!10}{\bf 85.5} \textcolor{green}{(+17.3)} & \cellcolor{gray!10}80.1 \textcolor{green}{(+59.8)} & \cellcolor{gray!10}84.2 \textcolor{green}{(+15.2)} & \cellcolor{gray!10}88.2 \textcolor{green}{(+4.3)} & \cellcolor{gray!10}85.1 \textcolor{green}{(+29.0)} \\
    \cline{2-9}
    & \multirow{2}{*}{SAM-ViT-H} & Box & 54.2 & 71.6 & 20.5 & 67.0 & 88.0 & 60.3 \\
    &  & \cellcolor{gray!10}PP & \cellcolor{gray!10}88.0 \textcolor{green}{(+33.8)} & \cellcolor{gray!10}{\bf 85.5} \textcolor{green}{(+13.9)} & \cellcolor{gray!10}{\bf 81.6} \textcolor{green}{(+61.1)} & \cellcolor{gray!10}{\bf 85.3} \textcolor{green}{(+18.3)} & \cellcolor{gray!10}{\bf 89.3} \textcolor{green}{(+01.3)} & \cellcolor{gray!10}{\bf 85.9} \textcolor{green}{(+25.6)}\\
    \bottomrule
  \end{tabularx}
  \caption{Performance of MedSAM\&SAM with conventional\&our prompts}
  \label{tab:ablation_pretrain}
  \vspace{-10pt}
\end{table*}

\subsection{Visualization of selective map effectiveness across different support images.} 
To evaluate the impact of Selective Map, we provided different support images for a single target image and observed how the Selective Map responded. MSE (Mean Square Error) values in the figure indicate the similarity between each support image and the target image. The heatmaps visualize the Selective Map’s attention, highlighting regions in the support image that the model deems relevant for segmentation. SI1 is identical to the target image (MSE = 0), while SI4 has the highest dissimilarity (MSE = 3010.1).

From a visualization aspect, the Selective Map consistently identifies semantically relevant regions, bypassing low-level appearance differences and capturing meaningful structural correspondences. From an output aspect, all support images (SI1-SI4) effectively assist segmentation, achieving highly stable Dice scores (minimum Avg. Dice = 94.42\%). The Selective Map identifies semantic correlations rather than relying solely on surface-level pixel alignment. However, given such stable and high performance, we further questioned the necessity of the support image. Therefore, we replaced it with a completely unrelated image (SI5, a retinal scan) that differs in modality, organ, and object structure. The Selective Map failed to provide meaningful guidance, thus the model's performance dropped sharply to 65.25\%, confirming that a relevant support image remains essential for accurate segmentation. However, the performance drop with SI5 is not a major concern in real-world applications, as clinicians naturally select relevant support images rather than arbitrarily using unrelated ones.


\subsection{Effect of encoder architectures in CSM.}
To further validate the structural rationality of our model, we replaced the Image Encoder in CSM with different architectures and evaluated their performance. For a fair comparison, Vision Mamba, Vision Transformer, and ResNet-50 were all trained without preloaded weights. The input channel of the first layer in each architecture was modified to four, allowing the encoder to extract features from both the original image and the mask. To ensure that the encoder output \(\mathbf{V}\) could be concatenated with the support feature map \(\mathbf{F}_{\text{sup}} \in \mathbb{R}^{K \times C \times H \times W}\) along the channel dimension, we aligned the spatial resolution of \(\mathbf{V}\) to match \(\mathbf{F}_{\text{sup}}\), both set to 1/16 of the original image size. For ResNet-50, we excluded layer 4, ensuring a 16 \(\times\) spatial downsampling relative to the input. For Vision Mamba and Vision Transformer, we set the patch size to 16, making the height and width of \(\mathbf{V} \in \mathbb{R}^{K \times C^{v} \times H \times W}\) 1/16 of the original image dimensions.  

For each dataset, we randomly selected 16 images for training. All three architectures were trained on the same dataset until convergence and evaluated using identical randomly sampled support pairs across the same test set. The per-object results are reported in \cref{tab:ablation_backbone}, where Vision Mamba achieves the best performance across all objects with the fewest trainable parameters.

\begin{figure*}[t]
  \centering
  \setlength{\abovecaptionskip}{3pt}
   \includegraphics[width=0.99\linewidth]{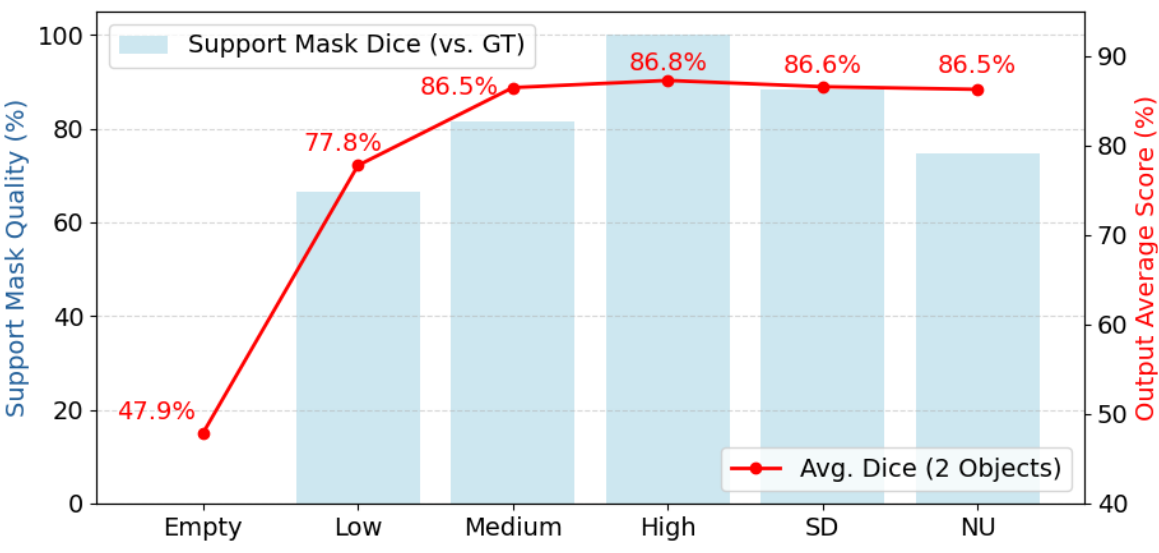}
   \caption{Line chart of model segmentation performance across varying prompt quality.}
   \label{fig:Diffmask}
   \vspace{-10pt}
\end{figure*}

\begin{figure*}[t]
  \centering
  \setlength{\abovecaptionskip}{3pt}
   \includegraphics[width=0.99\linewidth]{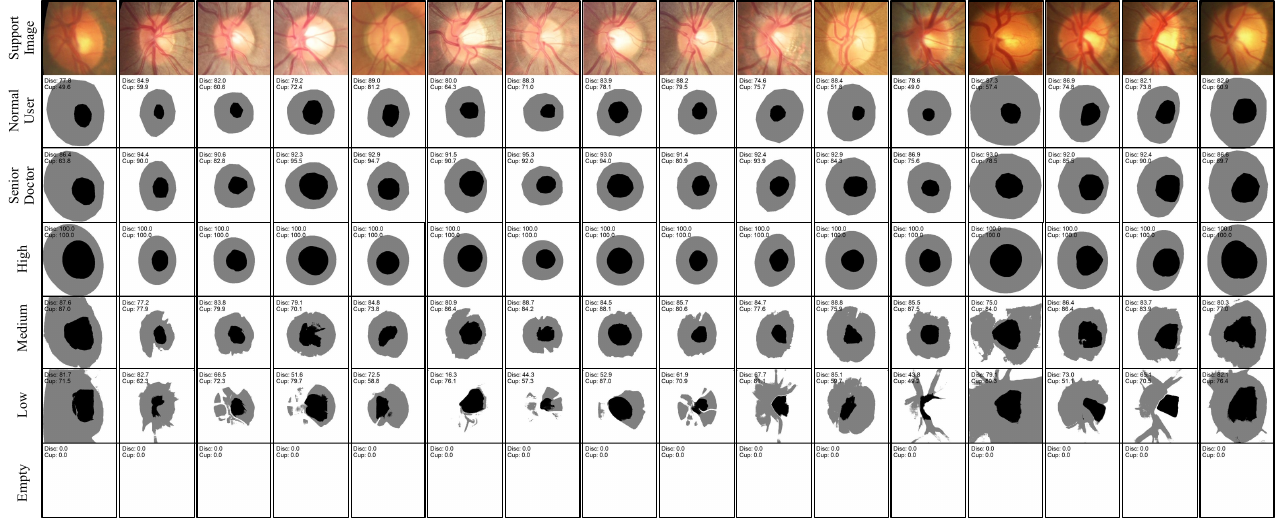}
   \caption{Six levels of support masks for 16 support images, with the gray area denoting the disc object and the black area indicating the cup object.}
   \label{fig:Diff_MASK_vis}
   \vspace{-15pt}
\end{figure*}

\subsection{Performance with different pretrained foundation models.}
We evaluated our method across different pretrained foundation models to assess its impact on both MedSAM and SAM when compared to their conventional box prompts. MedSAM, a SAM-based model trained on extensive medical datasets using box prompts, was tested alongside SAM models (ViT-B and ViT-H) using the same bounding box prompt (derived from the object’s minimum enclosing rectangle). For our approach, we randomly selected 16 training images, trained each model until convergence, and used randomly chosen support pairs as prompts for inference. The training dataset, test dataset and support pairs are same for all methods. From \cref{tab:ablation_pretrain}, we derive four key insights:

{\bf 1. Using box prompts, SAM-ViT-B, SAM-ViT-H, and MedSAM demonstrate a progressive improvement (56.1\% → 60.3\% → 74.5\%) in average Dice scores.} This trend reflects MedSAM's advantage in medical segmentation due to extensive domain-specific training and SAM-ViT-H's superior generalization from its larger parameter capacity compared to SAM-ViT-B.

{\bf 2. Despite MedSAM’s medical pretraining, our method further improves its segmentation performance on unseen datasets using only 16 support images.} The most notable improvement is on STARE-Vessel, where Dice increased by 39.3\%. This is because MedSAM struggles with vessel-like branching structures, as bounding box prompts can be ambiguous in such cases—a limitation noted in their paper~\cite{}. In contrast, our prompt leverages non-target data masks, inherently providing richer contextual information and enhancing precise segmentation.

{\bf 3. The extent of our method’s improvement on MedSAM correlates with its pretraining exposure to similar data.} According to MedSAM’s supplementary materials, its pretraining included FPA and REFUGE datasets. REFUGE, REFUGE2, and STARE all belong to the fundus imaging modality, but REFUGE2 introduces new data, while STARE has distinct segmentation objects. Thus, MedSAM's exposure to STARE, REFUGE2, and FPA datasets should increase in that order, while the performance gain from our method correspondingly decreases (+39.3\% → +7.1\%/2.5\% → -15.2\%/-10.6\%). The performance decline on the FPA dataset (-15.2\%/-10.6\%) is likely due to MedSAM's strong coupling of FPA data with the Box prompt during pretraining, causing conflicts when applying our different prompt strategy.

Moreover, the limitation of model improvement due to pretraining exposure is also show in our method enhancing MedSAM’s performance on STARE (20.1\% → 59.4\%) less than SAM-ViT-B’s (20.3\% → 80.1\%). Notably, MedSAM’s pretraining on fundus data exclusively focused on disc and cup segmentation, not vessels. One possible reason of limited improvement is that MedSAM’s have developed a prior focus on disc and cup segmentation during pretraining, leading it to instinctively segment optic disc/cup instead vessels. The prior focus is also reflected in MedSAM-Box scoring lower than SAM-Box (20.1\% vs. 20.3\%) on STARE and in qualitative results (main text Fig. 4, third row, third vs. second column).

These two observations indicate that MedSAM’s inherent pretraining biases influence its ability to adapt to new prompting strategies.

{\bf 4. Surprisingly, Our Method Boosts SAM-ViT-H Beyond MedSAM:}
Although our prompt improves MedSAM, it benefits SAM-ViT-H even more, enabling it to outperform MedSAM in segmentation (85.9\% v.s. 81.3\%). This contradicts our initial assumption that a medically pretrained model should be inherently better suited for medical segmentation. We attribute this to MedSAM’s rigid pretraining priors as we mentioned before, which hinder its flexibility in adapting to new prompts, similar to how students with prior but outdated knowledge may struggle more with conceptual shifts than those learning from scratch. Based on this, we suggest using MedSAM directly for datasets it was pretrained on but applying SAM+our method for unseen datasets, as the latter generalizes better.

Overall, our prompting strategy improves all three pretrained models. Notably, with our prompt, SAM-ViT-H achieves the best performance, even surpassing MedSAM on new datasets.

\subsection{Influence of prompt quality via support mask variations.} 
Since pre-annotation masks may vary in quality in real-world clinical practice, we analyzed our model’s performance under different prompt qualities using support mask variations. We first randomly selected 16 support images and generated masks at different quality levels. A completely empty mask with no object was categorized as $Empty$. $Low$ and $Medium$ masks were generated using SAM and MedSAM, both prompted with the object's minimum bounding box. The $High$ level mask was derived directly from the ground truth. To simulate variability in manual annotations, we incorporated human-labeled masks from annotators with different expertise levels. The $Senior Doctor$ masks were provided by an experienced physician, representing high-quality expert annotations. In contrast, the $Normal User$ masks were annotated by a non-clinician, simulating the challenges less experienced annotators may face in real-world scenarios. 

The line chart of model performance across varying prompt quality is shown in \cref{fig:Diffmask}, with support mask quality represented by blue bars indicating mask Dice. The red curve depicts precise Dice values, illustrating segmentation performance.

\cref{fig:Diff_MASK_vis} visualizes the 16 support images alongside their corresponding masks across six quality levels. The top-left corner of each subfigure shows the Dice score between the annotated object (disc or cup) and the ground truth. The average Dice scores for each mask level are as follows: $Empty$ (0.0\%), $Low$ (66.6\%), $Medium$ (81.6\%), $High$ (100\%), $Senior~Doctor$ (88.3\%), and $Normal~User$ (74.8\%). From the visual results, while the $Normal User$ mask has a lower Dice score than $Medium$, it still provides a clear indication of the target object. In contrast, at $Low$ and $Empty$ levels, the masks lack clarity in indicating the intended segmentation objects.

\subsection{Impact of training data size and support pair quantity in inference.} 
{\bf Experiment Details:} For each dataset, 16 samples were excluded to ensure a consistent test set across all experiments. We defined the 16 samples as \(\mathbf{Exclude\; set} = \{ (\mathbf{I}_{i}, \mathbf{M}_{i}) \}_{i=1}^{16}\). To evaluate the impact of training set size \( K \), we defined \(\mathbf{Train\; set} = \{ (\mathbf{I}_{i}, \mathbf{M}_{i}) \}_{i=1}^M\) with \( M \in \{2, 4, 8, 16\} \) and randomly selected training data from \(\mathbf{Exclude\; set}\), setting the number of support pairs to \( M-1 \) during training. For a given \( K \), we conducted repeated experiments by randomly selecting multiple diverse \(\mathbf{Support\; set} = \{ (\mathbf{I}_{i}, \mathbf{M}_{i}) \}_{i=1}^K\) from \(\mathbf{Exclude\; set}\) for each inference support size \( K \in \{1, 2, 4, 8\} \). For each \( M \) and \( K \), we conducted 100 repeated experiments, where in each trial, \( K \) support pairs were randomly selected from the \(\mathbf{Exclude\; set}\). The averaged Dice scores across three datasets and five objects were then used to generate the box plots and ridge plot (far right) in \cref{fig:ablation}, illustrating the average and variability of segmentation performance. 

The figure demonstrates a positive correlation between segmentation quality and \( K \). For each \( M \), the Dice score increases monotonically with \( K \) until \( K \) exceeds \( M-1 \), where a slight decline may occur. To further analyze the distribution density and concentration of results, we use a ridge plot to highlight experiments where \( K < M-1 \) for each \(M\). Peaks indicate where results are concentrated, while flatter regions suggest more fluctuated outcomes. The ridge plot reveals that when \( K=1 \), Dice scores are more widely distributed. However, as \( K \) increases, prediction variance across support pairs significantly decreases. For instance, in the top-right ridge plot (\( M=16 \)), results at \( K=8 \) (teal region) are more concentrated and form a sharper peak compared to \( K=1 \) (charcoal purple region), indicating that the results become more steady.
\begin{figure}[t]
  \centering
  \setlength{\abovecaptionskip}{1pt}
   \includegraphics[width=1\columnwidth]{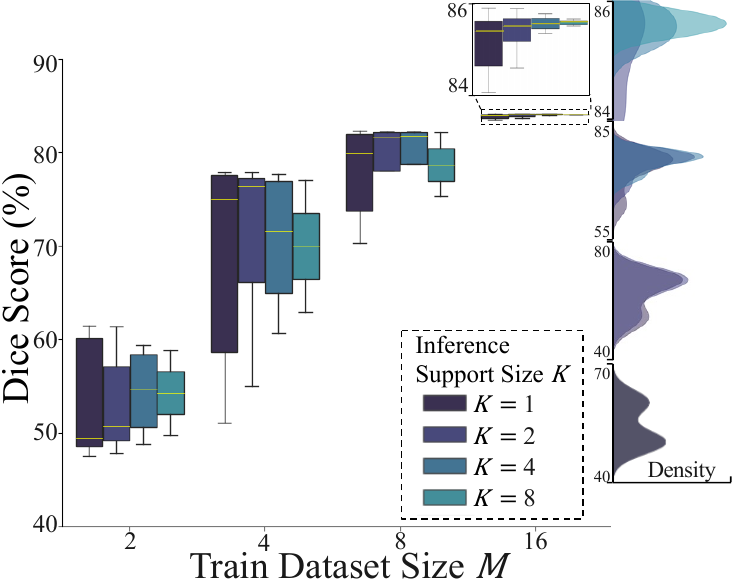}
   \caption{Enlarged box plots and ridge plots (rightmost column) of the ablation results .}
   \label{fig:ablation}
   \vspace{-20pt}
\end{figure}

\end{document}